\documentclass{elsart}

\usepackage{amssymb}
\usepackage{amsmath}
\usepackage{graphicx}
\usepackage{color}
\usepackage{multirow}
\usepackage{multicol}
\usepackage{longtable}
\usepackage{booktabs}

\newcommand{\be}{\begin{equation}}
\newcommand{\ee}{\end{equation}}
\newcommand{\ben}{\begin{eqnarray}}
\newcommand{\een}{\end{eqnarray}}
\newcommand{\lb}{\label}

\begin{document}

\begin{frontmatter}

\title{Hadronic states with both open charm and bottom in effective field
theory}

\author[UFBA]{L.M. Abreu}
\ead{luciano.abreu@ufba.br}

\corauth[cor2]{Corresponding author}

\address[UFBA]{Instituto de F\'{\i}sica, Universidade Federal da
Bahia, 40210-340, Salvador, BA, Brazil}

\begin{abstract}
  
We perform a field-theoretical study of possible deuteron-like molecules with 
both open charm and bottom, using the Heavy-Meson Effective Theory. In 
this approach, 
we analyze the parameter space of the coupling constants and discuss the 
formation of loosely-bound $D^{(*)}  B^{(*)} $-states. We estimate
their masses and other properties. 
  
\end{abstract}

\end{frontmatter}


\section{Introduction}

In the last decade we have witnessed a considerable progress in the hadron 
spectroscopy. In particular, experimental observations of unconventional hadron 
states have been reported by several experiments. These exotic 
 states, called $X \,Y \,Z$ states, exhibit unusual 
properties, as unexpected decay modes. For a review, see Refs. \cite{Brambilla,PDG}.  

From a theoretical point of view, a large amount of effort has been directed to
understand the structure of exotic states, and
several models have been proposed \cite{Brambilla2}. However, 
a natural interpretation that has 
been extensively used is to consider $X\,Y\,Z$ states as deuteron-like molecules of 
heavy-light mesons, due to the proximity of their masses to some hadronic thresholds.   
In this sense, these exotic hadrons might be yielded by the interaction of heavy 
hadrons, and interpreted as bound states if they are below the
threshold and in the first Riemann sheet 
of the scattering amplitude. 

It is worthy mentioning that the notion of molecular state with heavy-light 
hadrons was proposed about four decades ago, for the study of interaction between 
the charmed and anti-charmed mesons \cite{Voloshin}. Later, in subsequent decades 
this picture has been employed in different approaches, 
as in the quark-pion interaction framework for 
 analysis of several deuteron-like meson-meson bound states 
\cite{Tornqvist1,Tornqvist2}. But the discovery of exotic states has stimulated 
this concept, and it became a hot research topic of hadron physics. For example:
the $X(3872) $ state was proposed to be a loosely bound state of $D \bar{D}^{*}$
\cite{Tornqvist3,AlFiky,Braaten1,Dong1,Dong2,Lee,Braaten2,Nieves,Hidalgo,
Guo,Alberto}; another interesting 
 state is the $Z_b (10610)$,
considered as a $B \bar{B}^{*}$ molecule \cite{Guo,Bondar,Cleven}; in the case of  $Y 
(4260)$, it is interpreted as a $ D_0 \bar{D}^{*} $ 
\cite{Albuquerque,Ding,Wang}; and so on \cite{Brambilla2}.

In this scenario, although there is not yet experimental evidence, 
the issue about possible exotic states with masses in the region 
of $B_c$ sector (region of mass between the charmonium and bottomonium sectors)
 was also brought to light, and the interpretation of them 
as hadronic molecules with both open charm and open bottom has been raised 
\cite{Guo,Sun,Zhang,Albuquerque2,Li}. In particular, Ref. \cite{Guo} has used as 
guiding principle the heavy quark flavor symmetry. On the other hand, 
 the interaction between charmed and 
bottomed mesons has been investigated in Refs. \cite{Sun,Li} via the 
one-boson-exchange model, while Refs. \cite{Zhang,Albuquerque2} have explored some 
consequences of QCD sum rules.

We believe that there is still enough room for other contributions on this issue. 
In this work, we perform a field-theoretical study of possible deuteron-like molecules 
with both open charm and bottom, via the Heavy-Meson Effective Theory with 
four-body 
terms. An analysis is performed in some detail of the regions in parameter space  
in which the formation of loosely-bound $D^{(*)}  B^{(*)} $-states is allowed. 
We estimate their masses and other properties, and compare them with 
those existing in literature.  

This paper is organized as follows.  In Section~\ref{Formalism}, we present the 
formalism and the
method used to obtain the transition amplitudes and their solutions.
Section~\ref{Results} deals with the analysis of parameter space and the discussion of 
conditions for obtention of loosely bound $D^{(*)}  B^{(*)} $-states. 
We summarize the results and conclusions  in Section~\ref{Conclusions}. 
Some relevant tables are given in Appendix.

\section{Formalism}
\lb{Formalism}
\subsection{Heavy Meson Effective Lagrangian}

In order to investigate the bound states of $D^{(*)}  B^{(*)} $ with both charm 
and bottom, we must consider an effective theory that describes the interactions 
between heavy mesons, i.e. mesons containing a heavy quark $Q$.

Thus, we work with the effective theory known as Heavy Meson Effective Theory (HMET)
 \cite{AlFiky,Nieves,Manohar,Valderrama}. 
On this subject, we define the superfields: 
\begin{eqnarray}
\mathcal{H} _a ^{(Q)}= 
\left( \begin{array}{c}
H_a ^{(c)} \\
H_a ^{(b)} \end{array} \right);  \;\;\;\;\;
\mathcal{H}  ^{(\bar{Q}) a} =
\left( \begin{array}{cc}
H ^{(\bar{c}) a} , &
H ^{(\bar{b}) a} \end{array} \right), 
\label{H1}
\end{eqnarray}
where $Q = c,b$ is the index with respect to the heavy-quark flavor group 
$SU(2)_{HF}$, and 
\begin{eqnarray}
H_a ^{(Q)} & = & \left( \frac{1+ v_{\mu} \gamma ^{\mu} }{2}\right)\left( 
P _{a \mu}^{* (Q)} 
\gamma ^{\mu}  - P _{a }^{ (Q)}  \gamma ^{5} \right) ,\nonumber \\
H ^{(\bar{Q}) a} & = & \left( P _{ \mu}^{* (\bar{Q}) a} \gamma ^{\mu} - 
P ^{ (\bar{Q}) a}  \gamma^{5} \right) \left( \frac{1 - v_{\mu} 
\gamma ^{\mu}}{2}\right) .
\label{H2}
\end{eqnarray}
In Eq. (\ref{H2}), $v$ is the velocity parameter; $a$ is the triplet index of 
the light-quark flavor group  $SU(3)_V$; and
$P _{a }^{ (Q/ \bar{Q})}$ and $P _{a \mu}^{* (Q / \bar{Q})} $ are
the pseudoscalar and vector heavy-meson fields forming a $\mathbf{\bar{3}}$
representation of $SU(3)_V$:
\begin{eqnarray}
  P _{a }^{ (c)} & = & \left( D^{0},  D^{+}, D_s ^{+} \right), \nonumber \\
  P _{a }^{ (\bar{c})} & = & \left( \bar{D}^{0},  D^{-}, D_s ^{-} \right) ,
  \label{P1}
\end{eqnarray}
for the charmed meson field, and 
\begin{eqnarray}
  P _{a }^{ (b)} & = & \left( B^{-}, \bar{B}^{0}, \bar{B}_s ^{0} \right),\nonumber \\
  P _{a }^{ (\bar{b})} & = & \left(  B^{+}, B^{0}, B_s ^{0} \right) ,
  \label{P2}
\end{eqnarray}
for the bottomed meson field (and analogous expressions for the vector case). 

It is important to notice that the heavy vector meson fields obey the conditions: 
\begin{eqnarray}
 v \cdot  P _{a }^{* (Q)} & = & 0, \nonumber \\
  v \cdot  P ^{* (\bar{Q}) a} & = & 0 . 
\label{C1}  
 \end{eqnarray}
They define the three different polarizations of the heavy vector mesons.

The $\mathcal{H} _a ^{(Q)}$ and $ \mathcal{H} ^{(\bar{Q}) a }$ superfields 
transform under heavy-quark spin symmetry and $SU(3)_V$ light-quark flavor symmetry 
 as 
\begin{eqnarray}
  \mathcal{H} _a ^{(Q)} & \rightarrow & \mathcal{S} 
  \mathcal{H} _b ^{(Q)} 
 \mathcal{U}_{ b a }^{\dagger} , \nonumber \\
  \mathcal{H} ^{(\bar{Q}) a} & \rightarrow & \mathcal{U}^{ a b } 
  \mathcal{H}  ^{(\bar{Q}) b} 
 \mathcal{S}^{ \dagger} , 
 \label{H3}
  \end{eqnarray}
  where 
\begin{eqnarray}  
  \mathcal{S} = \left( \begin{array}{cc}
S ^{(c)}  & 0 \\
0         & S ^{(b)} \end{array} \right) \label{H4}
\end{eqnarray}
with $S ^{(c / b)}$ being the heavy-quark spin transformation, and 
\begin{eqnarray}
\mathcal{U} \in  SU(3)_V .
\label{H5}
\end{eqnarray}

Under heavy-quark flavor symmetry, the superfields transform as
\begin{eqnarray}
  \mathcal{H} _a ^{(Q) } & \rightarrow & U 
  \mathcal{H} _a ^{(Q) }  , \nonumber \\
  \mathcal{H} ^{(\bar{Q}) a} & \rightarrow &
  \mathcal{H}  ^{(\bar{Q}) a} U ^{ \dagger} , \label{H6}
  \end{eqnarray}
  where 
  \begin{eqnarray}
U \in  SU(2)_{HF} .
\nonumber
\end{eqnarray}
Notice that this symmetry relates only heavy mesons moving with the same velocity.

To construct invariant quantities under the symmetries discussed above, we need 
the hermitian conjugate fields: 
\begin{eqnarray}
\bar{\mathcal{H}}  ^{(Q) a} & = & \gamma ^0 \mathcal{H}_a  ^{(Q) \dagger } \gamma ^0 
= \left( \begin{array}{cc}
\bar{H} ^{(c) a} ,
\bar{H} ^{(b) a} \end{array} \right) \nonumber \\ \nonumber \\
\bar{\mathcal{H}}_a  ^{(\bar{Q}) }  & = & \gamma ^0 \mathcal{H}  ^{(Q) a \dagger }
 \gamma ^0
= \left( \begin{array}{c}
\bar{H}_a ^{(\bar{c})} \\
\bar{H}_a ^{(\bar{b})} \end{array} \right).
\label{H7}
\end{eqnarray}
They transform as  
  \begin{eqnarray}
  \bar{\mathcal{H}}  ^{(Q) a} & \rightarrow & \mathcal{U}^{ a b } 
  \bar{\mathcal{H}}  ^{(\bar{Q}) b} 
 \mathcal{S}^{ \dagger}, \nonumber \\
  \mathcal{H}_a ^{(\bar{Q}) } & \rightarrow &    
  \mathcal{S} 
  \mathcal{H} _b ^{(Q)} 
 \mathcal{U}_{ b a }^{\dagger} , \label{H8}
  \end{eqnarray}
and 
\begin{eqnarray}
   \bar{\mathcal{H}} ^{(Q) a} & \rightarrow &
  \bar{\mathcal{H}}  ^{(Q) a} U ^{ \dagger} , \nonumber \\
 \bar{\mathcal{H}} _a ^{(\bar{Q}) } & \rightarrow & U 
  \bar{\mathcal{H}} _a ^{(\bar{Q}) }  .\label{H9} 
 \end{eqnarray}

Now we are able to introduce the effective Lagrangian respecting heavy-quark 
spin, heavy-quark flavor and light-quark flavor symmetries. The Lagrangian at lowest 
order of the HMET can be written as
\begin{eqnarray}
\mathcal{L} = \mathcal{L}_2 + \mathcal{L}_4, 
\label{L1} 
\end{eqnarray} 
where the two-body piece is 
\begin{eqnarray}
\mathcal{L}_2 & = & - i \;\mathrm{Tr}\hspace{1pt} \left[ \bar{\mathcal{H}}  ^{(Q) b} v 
\cdot \mathcal{D} _{b}^{a}
 \;\mathcal{H}_{a}  ^{(Q) } \right]  - i \; \mathrm{Tr}\hspace{1pt} 
 \left[ \mathcal{H}  ^{(\bar{Q}) b}
  v \cdot \mathcal{D} _{b}^{a} \;
 \bar{\mathcal{H}} _{a} ^{(\bar{Q})  } \right] \nonumber \\
& & + i g \; \mathrm{Tr}\hspace{1pt} \left[
 \bar{\mathcal{H}}  ^{(Q) b} \mathcal{H}_{a}  ^{(Q) }
\gamma  ^{\mu} \gamma  ^{5}    \right] (\mathcal{A}_{\mu})_{b}^{a}
 + i g  \;\mathrm{Tr}\hspace{1pt}  \left[\mathcal{H}  ^{(\bar{Q}) b}  
\bar{\mathcal{H}}_{a}  ^{(\bar{Q})  } 
\gamma  ^{\mu} \gamma  ^{5}  \right] (\mathcal{A}_{\mu})_{b}^{a},
\label{L2} 
\end{eqnarray} 
with
\ben
(\mathcal{D} _{\mu })_{b}^{a}& = & \left[\partial _{\mu} + 
\frac{1}{2} \left( \xi ^{\dagger } 
\partial _{\mu} \xi + 
\xi \partial _{\mu} \xi  ^{\dagger }  \right) \right]_{b}^{a}, \;\;
(\mathcal{A}_{\mu})_{b}^{a} = \frac{1}{2} \left( \xi ^{\dagger } 
\partial _{\mu} \xi - 
\xi \partial _{\mu} \xi  ^{\dagger }  \right)_{b}^{a} ,\nonumber \\
\xi  & = &  e^{\frac{i}{f_{\pi}} M};  \;\;\;\;\;
M = \left( \begin{array}{ccc}
\frac{\pi ^{0}}{\sqrt{2}} + \frac{\eta}{\sqrt{6}} & \pi ^+  & K^+ \\
\pi ^- & -\frac{\pi ^{0}}{\sqrt{2}} + \frac{\eta}{\sqrt{6}}  & K^0 \\
K ^-   & \bar{K} ^0                         & - \frac{\eta}{\sqrt{6}} 
 \end{array} \right).
\label{M} 
\een  
The four-body interaction piece reads
\begin{eqnarray}
\mathcal{L}_4 & = & - \frac{D_1}{4} \mathrm{Tr}\hspace{1pt}
\left[ \bar{\mathcal{H}}  ^{(Q) a} 
 \mathcal{H}_{a} ^{(Q) } \gamma  ^{\mu} \right] \mathrm{Tr}\hspace{1pt} \left[
 \mathcal{H}  ^{(\bar{Q} ) a}
 \;\bar{\mathcal{H}}_{a}  ^{(\bar{Q}) } \gamma  _{\mu} \right]  \nonumber \\
& & - \frac{D_2}{4} \mathrm{Tr}\hspace{1pt}\left[ \bar{\mathcal{H}}  ^{(Q) a} 
 \mathcal{H}_{a}  ^{(Q) } \gamma  ^{\mu} \gamma  ^{5} \right] \mathrm{Tr}\hspace{1pt} \left[
 \mathcal{H} ^{(\bar{Q}) a}
 \;\bar{\mathcal{H}}_{a}  ^{(\bar{Q})  } \gamma  _{\mu} \gamma  ^{5}
  \right]   \nonumber \\
& &   - \frac{E_1}{4} \mathrm{Tr}\hspace{1pt} \left[ \bar{\mathcal{H}}  ^{(Q) a}
 (\lambda ^{A})_{a}^{b}
 \mathcal{H} _{b} ^{(Q) } \gamma  ^{\mu} \right] \mathrm{Tr}\hspace{1pt} \left[
 \mathcal{H} ^{(\bar{Q}) a}
  (\lambda _{A})_{a}^{b} \;\bar{\mathcal{H}}_{b}  ^{(\bar{Q}) } \gamma  _{\mu} \right] 
   \nonumber \\
& &  - \frac{E_2}{4} \mathrm{Tr}\hspace{1pt} \left[ \bar{\mathcal{H}}  ^{(Q) a} 
 (\lambda ^{A})_{a}^{b}
 \mathcal{H}_{b}  ^{(Q) } \gamma  ^{\mu} \gamma  ^{5} \right]  \mathrm{Tr}\hspace{1pt}
 \left[ \mathcal{H}  ^{(\bar{Q}) a} (\lambda _{A})_{a}^{b}
 \;\bar{\mathcal{H}}_{b}  ^{(\bar{Q})  } \gamma  _{\mu} \gamma  ^{5}
  \right]  ,  
  \label{L4}
  \end{eqnarray}
where $\lambda _{A}$ are the Gell-Mann matrices. 

In Eq. (\ref{L1}), we must consider 
in the products of the superfields $\mathcal{H}$ the different 
 heavy-quark flavor and light-quark flavor spaces, as pointed in the definitions 
 in Eqs. (\ref{H1}) and (\ref{H7}). So these products must be performed properly.

It is worthy mentioning that there are other Lorentz Structures at leading order.  
However, as remarked in Ref. \cite{Liu} these other contact terms are not 
independent; they are linear combinations of terms in Eq. (\ref{L4}). Thus, we 
will omit them.

We work in the leading order in the $1/m_Q$ expansion. Thus,  
relativistic effects are suppressed and two-heavy meson system can be described in 
the non-relativistic version of the theory. In this sense, it is convenient to adopt 
the velocity parameter as $v = \left( 1, \vec{0} \right)$, and employ the following 
normalization
\cite{Manohar,Valderrama}: : 
\be 
  \sqrt{2}  P _{a }^{(* \mu)} \rightarrow  P _{a }^{(* \mu) }, 
\label{norm}
\ee
It can be remarked that the choice above of $v$ makes the component 
$\mu =0$ of the vector meson irrelevant. Therefore, we work only with 
the euclidian part of the vector meson fields henceforth. 

It is also interesting to analyze the power counting in this scheme: with respect 
to the heavy quantity, the heavy-quark 
mass $m_Q$, 
 we see that the kinetic term in Eq.~(\ref{L2}) 
gives the scaling for the Lagrangian density as 
$\mathcal{L} \sim m_Q ^{0}$ \cite{Manohar,Casalbuoni}. So, $\mathcal{L}$ 
does not scale with $m_Q$. 
Therefore, we obtain 
the following scaling for the meson fields: $\mathcal{H} \sim
m_Q ^{0}$ and $M \sim m_Q ^{0}$. Thus, the scaling analysis of the terms in 
Eq.~(\ref{L4}) yields: $D_i, E_i \sim m_Q ^{0} $. Hence, the terms of 
$\mathcal{L}_2$ and $\mathcal{L}_4$ in Eq.~(\ref{L1}) are leading order $O(1)$ in the
$1/m_Q$ expansion.  In addition, the mentioned terms are also the leading order 
in the chiral expansion; corrections to lowest order come from 
higher derivative or mass terms and from loop diagrams \cite{Manohar,Casalbuoni}.

At this point we must discuss some questions of our approach. In the scenario of heavy 
hadronic molecules, pion-exchange effects are in general perturbative over the 
expected range of applicability of HMET and are suppressed, as pointed in Refs. 
\cite{Nieves,Guo,Valderrama}. This situation is in contrast to two-nucleon 
systems (usually used as similar systems to heavy-meson molecules), in which the 
leading order potential of Chiral Perturbation Theory includes one-pion exchange 
interaction \cite{Birse,Epelbaum}. As a consequence, at lowest order the HMET 
can be considered as a contact-range theory, taking into account the proper 
range of binding energies. Thus, following these findings we explore the 
leading-order potential of HMET only with contact interactions present in Eq. 
(\ref{L4}), and investigate the region where the pion-exchange contribution 
is not relevant.

Hence, considering the discussion above, we can perform 
 the expansion of the $\mathcal{H}$-fields in the heavy-quark limit. After some 
 manipulations, the four-body interaction terms in Eq. (\ref{L4}) read
\begin{eqnarray}
\mathcal{L}_4 & = & D_1 \left( P ^{* (Q) a \dagger } \cdot P_a ^{* (Q)  }  
+ P ^{(Q) a \dagger } P_a ^{ (Q) } \right)  
\left( P ^{* (\bar{Q}) a^{\prime}} \cdot P_{a^{\prime}}
 ^{* (\bar{Q}) \dagger }  
+ P ^{(\bar{Q}) a^{\prime}} P_{a^{\prime}} ^{ (\bar{Q}) \dagger } \right)
\nonumber \\
& & - D_2 \left( P ^{* (Q)a \dagger } \times  P_a ^{* (Q)  }  \right) 
\cdot \left(  P ^{* (\bar{Q}) a^{\prime}} \times P_{a^{\prime}} ^{* (\bar{Q}) 
\dagger } \right)  \nonumber \\
& & - i D_2 \left[ \left( P ^{* (Q)a \dagger } \times  P_a ^{* (Q)  }  \right) 
\cdot  \left( P ^{* (\bar{Q}) a^{\prime}} P_{a^{\prime}} ^{ (\bar{Q}) \dagger }  
+ P ^{(\bar{Q}) a^{\prime}} P_{a^{\prime}} ^{ * (\bar{Q}) \dagger } \right)
\right. \nonumber \\
& & + \left. \left( P ^{* (Q) a \dagger } P_{a} ^{ (Q) }  
+ P ^{(Q)a \dagger } P_{a} ^{ * (Q)  } \right) \cdot 
\left( P ^{* (\bar{Q}) a^{\prime}} \times  P_{a^{\prime}} ^{* (\bar{Q})  \dagger }
 \right) \right] \nonumber \\
& &  - D_2 \left( P ^{* (Q)a \dagger } P_a ^{ (Q)  }  
+ P ^{(Q) a \dagger } P_a ^{ * (Q) } \right) \cdot
 \left( P ^{* (\bar{Q}) a^{\prime}} P_{a^{\prime}}
 ^{(\bar{Q}) \dagger }  
+ P ^{(\bar{Q})a^{\prime} } P_{a^{\prime}} ^{ * (\bar{Q}) \dagger } \right)
\nonumber \\
& & + \left\{ \left( D_i \rightarrow E_i \right) \left[  P ^{(*) a}\circ P_a  ^{(*)}
 \rightarrow P  ^{(*) a}\circ 
\left(\lambda ^{A} \right)_{a}^{b}  P_b  ^{(*)}  \right] \right. \nonumber \\
& & \diamond \left. \left[  P ^{(*) a^{\prime}}\circ P_{a^{\prime}}  ^{(*)}
 \rightarrow 
P  ^{(*) a^{\prime}}\circ \left(\lambda _{A} \right)_{a^{\prime}}^{ b^{\prime}} 
 P_{b^{\prime}}  ^{(*)} \right]  \right\}.
 \label{L4mod}
 \end{eqnarray}
In this equation, the polarization of the vector mesons $P_a ^{ *}$ and 
the sum over the two heavy-quark flavors are
considered implicit. 
The last two lines in Eq. (\ref{L4mod}) mean the addition of equivalent terms 
to the ones expressed in lines above, with the replacement of respective bilinear 
$P^{(*)}P^{(*)}$-forms  (diagonal in light-quark $SU(3)_V$ space) by other ones carrying 
Gell-Mann matrices. Thus, we see that the interaction strength is described by 
four parameters: $D_1, D_2, E_1 $ and $ E_2$. Next, we will analyze the influence of 
these parameters on the existence of 
 bound states and their resulting binding energies.    

\subsection{Transition Amplitudes}

We are interested in the analysis of the scattering
\be 
 D^{(*)}(1)  B ^{(*)} (2) \rightarrow D^{(*)}(3)  B ^{(*)} (4). 
\label{scat}
\ee
Then, we can use the Breit approximation in order to relate the non-relativistic 
interaction potential, $V$, and the scattering amplitude $ i \mathcal{M}
(D^{(*)} B ^{(*)} \rightarrow D^{(*)} B ^{(*)} )$:
\ben 
V(\vec{p}) = - \frac{1}{\sqrt{\Pi _i 2 m_i \Pi _f 2 m_f}} \mathcal{M}(D^{(*)} B ^{(*)}
 \rightarrow D^{(*)} B ^{(*)} )
\label{rel1} 
\een
where $m_i $ and $m_f$ are the masses of initial and final states, and $\vec{p}$
is the momentum exchanged between the particles in Center-of-Mass frame. 

Following Ref. \cite{Valderrama}, it is convenient to 
 categorize the $D^{(*)} B ^{(*)}$-states into four groups: $D B, \,
D^{*} B , \, D B ^{*}$ and $D^{*} B ^{*}$. Then, by considering the 
four-body Lagrangian $\mathcal{L}_4$ in Eq. (\ref{L4mod}), it is possible to obtain  
the scattering amplitude at tree-level approximation, 
yielding the effective potential $V $ in the basis of states 
$\mathcal{B}
\equiv \left\{  | D B \rangle,
 \, | D^{*} B \rangle, \, | D B^{*}\rangle, \, 
| D^{*} B^{*} \rangle  \right\} $.
 The result is shown in Table 
\ref{table1}.

\begin{table}[htdp]
  \caption{Terms of interaction potential $V (\vec{p})$ in the basis 
$\mathcal{B} 
\equiv \left\{  | D B \rangle, \,
 | D^{*} B \rangle, \, | D B^{*}\rangle, \, 
| D^{*} B^{*} \rangle  \right\} $. $\vec{\varepsilon}_i$ means the polarization 
of incoming or outgoing vector heavy meson; $\vec{S}_i$ is the spin-1 operator, 
whose matrix elements are equivalent to the vector product of polarizations; and 
$C_i  =  D_i + E_i \lambda _A \lambda ^A$.  
}
\begin{center}
\begin{tabular}{|c|c|c|c|c|} \hline \hline
 & $D B$ & $D^{*} B$ &  $D B ^{*}$  &  $D^{*} B ^{*}$ \\  \hline \hline
$D B$ & $C_1$  &  0    &    0   & 
$- C_2 \vec{\varepsilon}_1 \cdot \vec{\varepsilon}_2 $\\
$D^{*} B$ & 0  & $C_1 \vec{\varepsilon}_3 ^{*} \cdot \vec{\varepsilon}_1$ 
   & $- C_2 \vec{\varepsilon}_3 ^{*} \cdot \vec{\varepsilon}_2$
& $- C_2 \vec{\varepsilon}_2 \cdot \vec{S}_1$ \\
$ D B ^{*} $ & 0  &  $- C_2 \vec{\varepsilon}_4 ^{*} \cdot \vec{\varepsilon}_1 $  
 & $ C_1 \vec{\varepsilon}_4 ^{*} \cdot \vec{\varepsilon}_2 $
 & $ C_2 \vec{\varepsilon}_1 \cdot \vec{S}_2$ \\
 $D^{*} B ^{*}$ & $- C_2 \vec{\varepsilon}_3  ^{*}\cdot \vec{\varepsilon}_4 ^{*}$ 
&  $- C_2 \vec{\varepsilon}_4 ^{*} \cdot \vec{S}_1$
& $C_2 \vec{\varepsilon}_3 ^{*} \cdot \vec{S}_2$ 
& $C_1 \vec{\varepsilon}_3 ^{*} \cdot \vec{\varepsilon}_1
\vec{\varepsilon}_4 ^{*} \cdot \vec{\varepsilon}_2 + C_2\vec{S}_1 \cdot 
\vec{S}_2$ \\
 \hline
\end{tabular}
\end{center}
\label{table1}
\end{table}

In Table \ref{table1}, we have used the notation \cite{Valderrama}: 
\ben
C_i & = & D_i + E_i \lambda _A \lambda ^A ;\;\; i = 1,2 ; \nonumber  \\
\vec{S}_1 & \equiv &  (\vec{\varepsilon}_3 ^{*}  \times \vec{\varepsilon}_1); \nonumber \\  
\vec{S}_2 & \equiv &  (\vec{\varepsilon}_4 ^{*}  \times \vec{\varepsilon}_2).
\label{CS} 
\een

At this point, it is convenient to explicit the relation between the  $SU(3)_V$
light-quark flavor and particle basis. In this sense, each state of the 
$\mathcal{B} $-basis is composed of nine light-quark flavor states, 
i. e. one octet and one singlet. Specifically, there are  two isosinglets ($I=0,
S=0 $), 
\ben 
| a_{s1} \rangle &  \equiv &  \frac{1}{\sqrt{2}} 
\left[ | D^{(*) 0} B^{(*)+} \rangle + 
| D^{(*) +} B^{(*) 0} \rangle \right] , \nonumber \\
| a_{s2} \rangle & \equiv  &  | D _s ^{(*) +} B _s ^{(*)0} \rangle ; \label{iso}
\een
one isotriplet ($I=1, S=0 $), 
\ben 
| a_{t} \rangle &  \equiv & 
\left\{ 
| D^{(*) 0} B^{(*)0} \rangle , 
 \frac{1}{\sqrt{2}} \left[ | D^{(*) 0} B^{(*)+} \rangle -
| D^{(*) +} B^{(*) 0} \rangle \right] , 
| D^{(*) +} B^{(*)+} \rangle \right\};
\label{tri}
\een 
%
and two isodoublets ($I=\frac{1}{2}, S= \pm 1 $),  
\ben 
| a_{d1} \rangle &  \equiv &  \left\{ | D _s ^{(*) +} B^{(*)0} \rangle ,
| D _s ^{(*) +} B^{(*)+} \rangle \right\} , \nonumber \\ 
| a_{d2} \rangle &  \equiv &  \left\{ 
| D  ^{(*) +} B _s ^{(*)0} \rangle ,
| D  ^{(*) 0} B _s ^{(*)0} \rangle \right\} .
\label{doublet} 
\een

In this scenario, the coupling constants $C_1$ and $C_2$ are given by the following 
expressions with respect to the specific channels of light-quark 
flavor $SU(3)_V$ basis:
\ben 
| a_{s1} \rangle & : &  C_i = 2 D_i + \frac{20}{3} E_i, \nonumber \\
| a_{s2} \rangle  & : &  C_i = 2 D_i + \frac{8}{3} E_i, \nonumber \\
| a_{t} \rangle  & : &  C_i = 2 D_i - \frac{4}{3} E_i ,\nonumber \\
| a_{d1} \rangle  & : &  C_i = 2 D_i - \frac{4}{3} E_i , \nonumber \\ 
| a_{d2} \rangle   & : &  C_i = 2 D_i - \frac{4}{3} E_i .
\label{C2} 
\een

In order to obtain dynamically generated poles in the amplitudes, we work 
with transition amplitudes satisfying the Lippmann-Schwinger equation
\ben
T ^{(\alpha \beta)} = V ^{(\alpha \beta)} + \int \frac{d^{4}q}{(2 \pi )^{4}} 
V^{(\alpha \gamma)} \,G \, T ^{(\gamma \beta)}, 
\label{LS1}
\een
where $\alpha, \beta, \gamma = | H\bar{H}_{\xi} \rangle $, with  
 $H\bar{H} = D B, \,
D^{*} B , \, D B ^{*}, D^{*} B ^{*}$ and $  \xi = s1, s2, t , d1, d2 $
 representing each channel associated to the 
 $ \mathcal{B}$  and
 light-quark flavor $SU(3)_V$  bases, respectively. Also,  
\ben 
G \equiv
\frac{1}{\frac{\vec{p}^{2}}{2 m_{D^{(*)}} } + 
q_0 - \frac{\vec{q}^{2}}{2 m_{D^{(*)}} } + i \epsilon} \;\;
\frac{1}{\frac{\vec{p}^{2}}{2 m_{B^{(*)}} } + 
q_0 - \frac{\vec{q}^{2}}{2 m_{B^{(*)}}}  + i \epsilon }. 
\label{G1}
\een
Notice that for vector mesons, we must perform the replacement
  $ G \rightarrow G^{\mu \nu } $.

The solutions of Lippmann-Schwinger equation for a 
specific channel in the present case have the 
form
\ben 
T ^{ (\alpha) } = \frac{V ^{ (\alpha) } }{1 - V ^{ (\alpha) } G ^{ (\alpha) } }. 
\label{LS2}
\een


In the analysis of pole structure of Lippmann-Schwinger equation, resonances are 
understood as the poles located 
in the fourth quadrant of the momentum complex plane (in the second Riemann sheet), 
while bound states are below the threshold (in the first Riemann sheet). Thus, 
since here we are interested 
on bound-state solutions, the use of residue theorem and dimensional regularization
in Eq. (\ref{LS2}) yield \cite{AlFiky} 
\ben 
T ^{ (\alpha) } = \frac{\tilde{V} ^{ (\alpha) } }{1 + \frac{i}{8 \pi} 
\mu |\vec{p}| \;\tilde{V} ^{ (\alpha) } } ;
\label{CS3}
\een
where $\tilde{V} ^{ (\alpha) }$ is the renormalized potential (renormalized contact interaction), 
and  $\mu  $ is reduced mass of $D^{(*)} B^{(*)}$ system. In the renormalization 
procedure above, $\tilde{V} ^{ (\alpha) }$ is a quantity dependent of 
renormalization scheme, since the bare coupling constants are adjusted in order 
to absorb the ultra-violet divergences (although this divergence does not explicitly 
manifest in the dimensional regularization above) contained in 
the Lippmann-Schwinger equation. 

 Therefore, other quantities 
can be obtained from Eq. (\ref{CS3}), as the binding 
energy, 
\ben 
E_b ^{ (\alpha) } = \frac{32 \pi^{2} }{\left( \tilde{V} ^{ (\alpha) } \right)^2 
\mu ^3 }, 
\label{BE1}
\een
and scattering length, 
\ben 
a_s ^{ (\alpha) } = \frac{ \mu \tilde{V} ^{ (\alpha)  }  }
{ 8 \pi } .
\label{SL}
\een
 
We remark that the contact interaction in Eq. (\ref{CS3}), (\ref{BE1}) and 
(\ref{SL}), implicit in $\tilde{V}$,
 must be renormalized. However, to simplify the notation we continue to denote the 
renormalized parameters as $D_1, D_2, E_1 $ and $ E_2$. 

It is also worthy mentioning that despite the renormalization dependence of 
 $\tilde{V} ^{ (\alpha)}$, quantities such as binding energy and scattering 
length are observables, and therefore renormalization-independent.

\section{Results}
\lb{Results}

After the obtention of a general solution of transition amplitude, 
now we analyze the possibility of bound states of $D^{(*)} B^{(*)}$ systems in the 
context discussed above. As remarked, we restrict our analysis to the region of 
relevance of contact-range interaction, that is the region where the pion-exchange 
contribution is not relevant. In this sense, we study bound states which obey 
the requirement $a_S \gtrsim 3 \lambda _{\pi}$, where $\lambda _{\pi } = 1/ m_{\pi}
\sim 1 $ fm is the pion 
Compton wavelength. 

It is relevant to remark that although the renormalization procedure should be 
performed in each sector (with different coupling strengths for each sector), we 
work here in a specific renormalization scheme, in which the relation between 
the bare coupling constants is conserved after renormalization. This approach 
allows to relate the results for different sectors, taking into account the fact 
that there is not yet experimental evidence of hadronic molecules with both open 
charm and open bottom.  

In this sense, using Eqs. (\ref{CS3}), (\ref{BE1}) and (\ref{SL}) we study the 
mass, binding energy and scattering length of $S$-wave bound states as 
functions of interaction strength.

\subsection{$DB({}^1 S_0), D^{*}B({}^3 S_1)$ and $D B^{*} ({}^3 S_1)$ systems}

We start by analyzing the $| DB({}^1 S_0) \rangle , | D^{*}B({}^3 S_1) \rangle$ 
and $ | D B^{*} ({}^3 S_1) \rangle $ states. As shown in Table \ref{table1}, they 
depend only on the parameter  
$C_1= D_1 + E_1 \lambda _A \lambda ^A $. Thus, the $(D_1,E_1)$-parameter 
space can be explored. 

\begin{figure}[htbp] 
	\centering
	\includegraphics[width=0.8\textwidth]{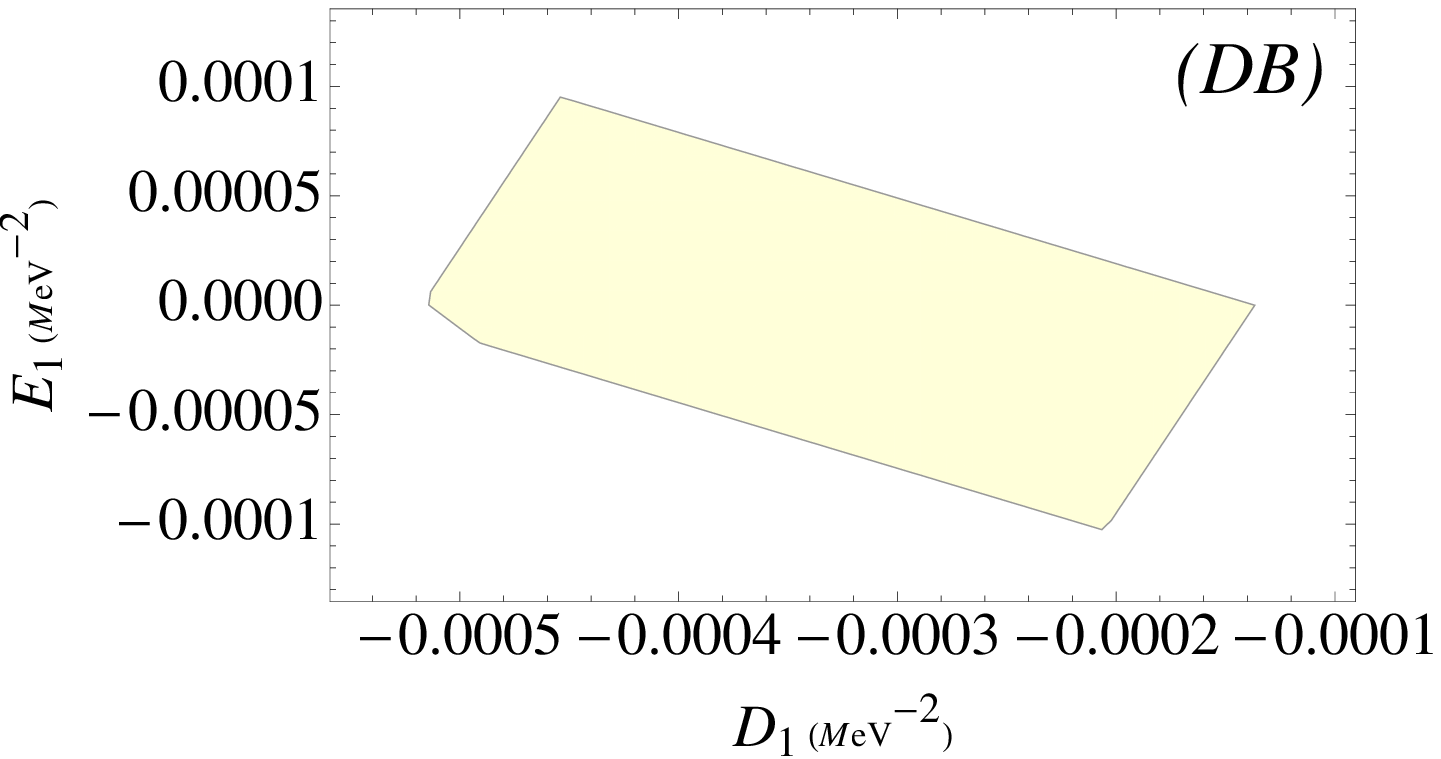} \\
	\includegraphics[width=0.8\textwidth]{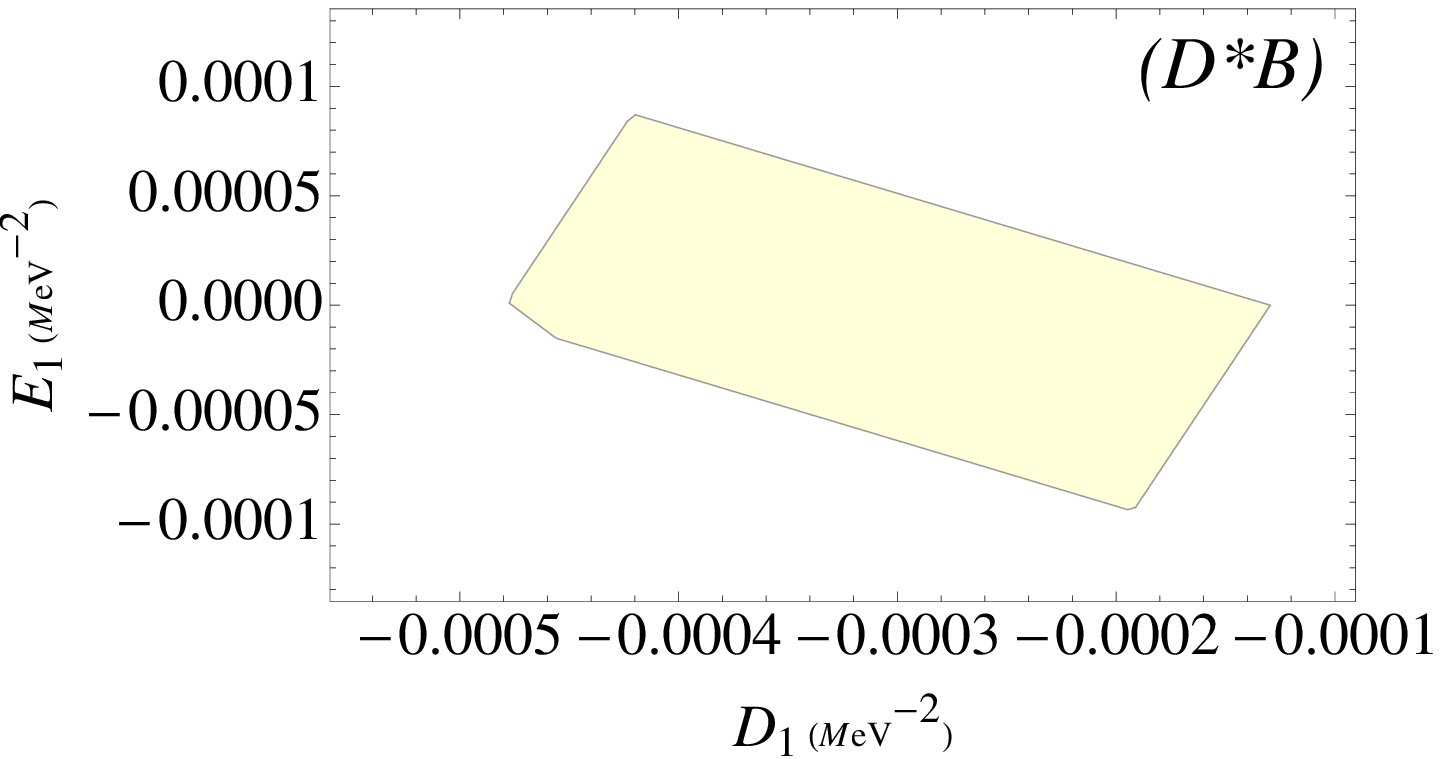} \\
	\includegraphics[width=0.8\textwidth]{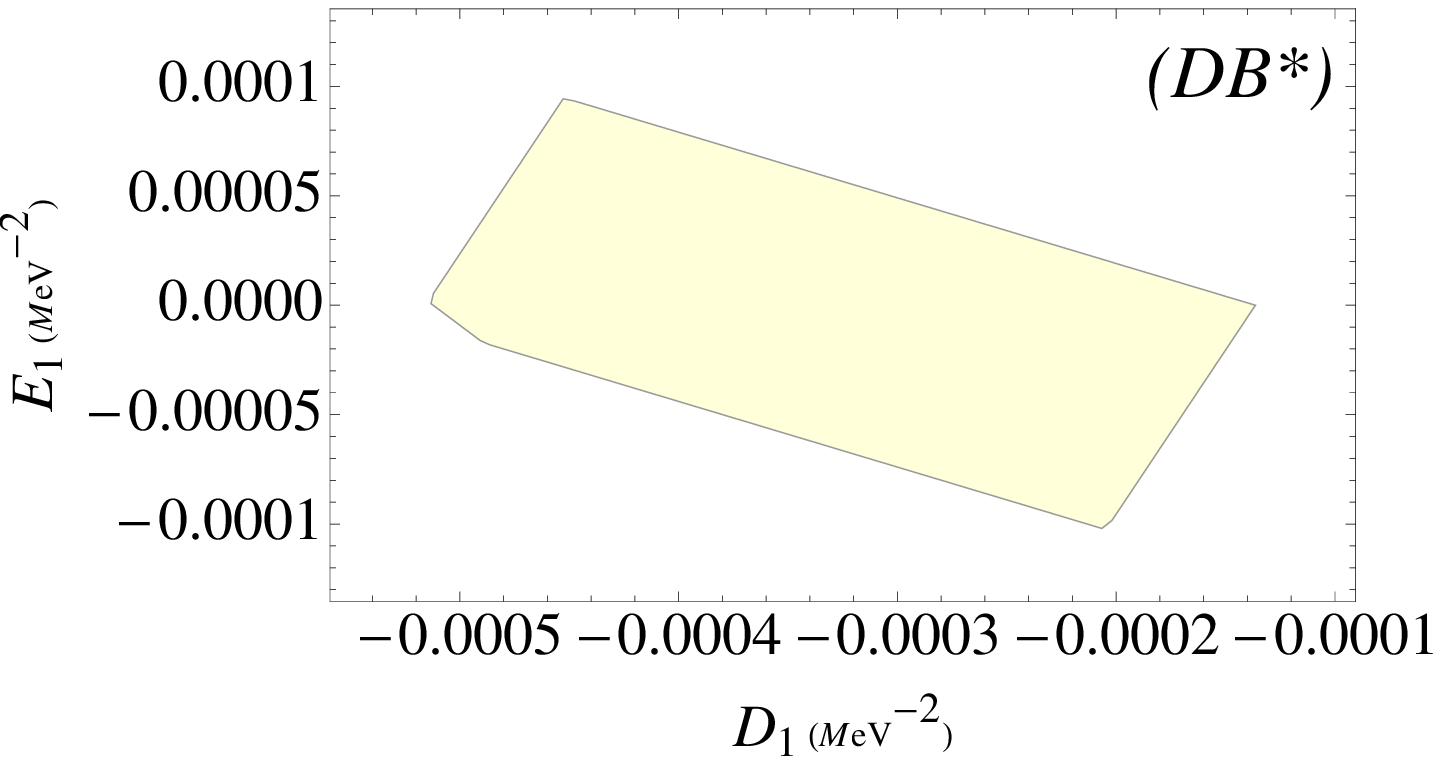}
\caption{$(D_1,E_1)$-parameter 
space; light shaded areas represent the regions in which the parameters acquire 
values that allow bound states 
for the nine states of $SU(3)_V$ light-quark flavor 
basis shown in Eq. (\ref{C2}),  with binding energy greater 
than 0.1 MeV and obeying the condition $a_S \gtrsim 3 \lambda _{\pi}$. 
Upper, middle and lower panels represent the $DB({}^1 S_0)_{\xi}, D^{*}
B({}^3 S_1)_{\xi}$ and $D B^{*} ({}^3 S_1)_{\xi}$ systems, respectively 
( $\xi = s1, s2, t, d1, d2$ ). }
	\label{FIGX1}
\end{figure}

In Fig. \ref{FIGX1} the light shaded areas indicate the 
intersection region in which bound states are obtained with binding energy greater 
than 0.1 MeV and obeying the condition $a_S \gtrsim 3 \lambda _{\pi}$. 
In this region the relevant 
parameters acquire values that allow loosely bound states 
for the nine states of $ SU(3)_V$ light-quark flavor basis
explicited in 
Eq. (\ref{C2}). 
 	
Using this analysis of the $(D_1,E_1)$-parameter 
space, we study the mass, binding energy and scattering length of the 
$| DB({}^1 S_0) \rangle , | D^{*}B({}^3 S_1)  \rangle$ and 
$ | D B^{*} ({}^3 S_1)\rangle $ 
states for different values of the parameters $(D_1,E_1)$. These values are 
chosen in a such way that their magnitudes are smaller, similar or greater than 
the ones delimited 
by the intersection region of last panel in Fig. \ref{FIGX1}:
\begin{description}
\item[$(i)$]  $D_1=-0.000015\; \mathrm{MeV}^{-2},\; E_1=-0.00001 \; 
\mathrm{MeV}^{-2}$; 
\item[$(ii)$] $D_1=-0.00005\; \mathrm{MeV}^{-2},\;   E_1=-0.00002 
\; \mathrm{MeV}^{-2}$; 
\item[$(iii)$]  $D_1=-0.0003 \; \mathrm{MeV}^{-2}, \;   E_1=-0.00005 \;
\; \mathrm{MeV}^{-2}$; 
\item[$(iv)$]  $D_1=-0.001 \; \mathrm{MeV}^{-2}, \;    E_1=-0.0001 \; 
\mathrm{MeV}^{-2}$. 
\end{description}
In general, the choice $(iii) $ works with sufficient 
magnitudes of $(D_1,E_1)$ to be found on the  
 the light shaded area of panels in Fig. \ref{FIGX1}, while $(i)$ and $(ii)$ [$(iv)$] 
 give magnitudes smaller [greater] than the ones mentioned above 
 and are not inside the mentioned light-shaded region. 
 
The results are shown in Tables \ref{table2},  \ref{table3} and \ref{table4}
of Appendix \ref{AppA}. The errors in our predicted quantities are obtained 
taking into account the violations of heavy quark spin symmetry 
due to the finite heavy quark mass and the breaking of light flavor symmetry. These 
effects are the most relevant high-order corrections to the heavy meson contact 
interactions. For the first of two effects, we expect a relative uncertainty of the 
order of $\Lambda _{QCD}/m_Q$ 
of the values of $\tilde{V} ^{ (\alpha) }$ in heavy-quark limit. Taking 
$\Lambda _{QCD} \sim 200 $ MeV and the quark charm mass $m_c \sim 1.5$ GeV
\cite{Hidalgo}, this estimated error is of 15\% in leading order contact 
interactions.  The second deviation, due to $SU(3)$-breaking effects, can be 
estimated from the ratio of kaon and pion decay constants, 
$f_K / f_{\pi}\sim 1.2$, giving a relative error of 
20\% in molecules containing strange quarks. Thus, the total error in 
$\tilde{V} ^{ (\alpha) }$ is calculated 
by adding the partial error in quadrature. After that it is possible to estimate 
the uncertainties in binding energy and in scattering length given by  Eqs. 
(\ref{BE1}) and (\ref{SL}), respectively.

 In all situations displayed in Tables \ref{table2},  \ref{table3} and \ref{table4},
 the 
bound states do not seem possible for the choice $(iv)$. On the other hand, we find 
loosely-bound state solutions  for all nine $ SU(3)_V$
 light-quark flavor 
states ($\xi = s1, s2, t, d1, d2$) in the situation $(iii)$, while in the 
case of choice (ii) only for the isosinglet $s1$.

Besides, the results suggest that for both spin-0 and spin-1 systems, 
the $t,d1,d2$ channels present greater binding energies than $s1, s2$  
in a specific choice of parameters.

\subsection{$D^{*} B^{*}({}^1 S_0), D^{*} B^{*}({}^3 S_1)$ and 
$D^{*} B^{*}({}^5 S_2)$ systems}

For the $D^{*} B^{*}({}^1 S_0), D^{*} B^{*}({}^3 S_1)$ and 
$D^{*} B^{*}({}^5 S_2)$ systems, the parameter space is richer, 
since the transition amplitudes depend on the constants  
$C_1$ and $C_2$, i.e. on $D_1, E_1, D_2  $ and $E_2$, according to corresponding 
 channel of effective potential shown in Table \ref{table1}.

\begin{figure}[htbp] 
	\centering
	\includegraphics[width=0.8\textwidth]{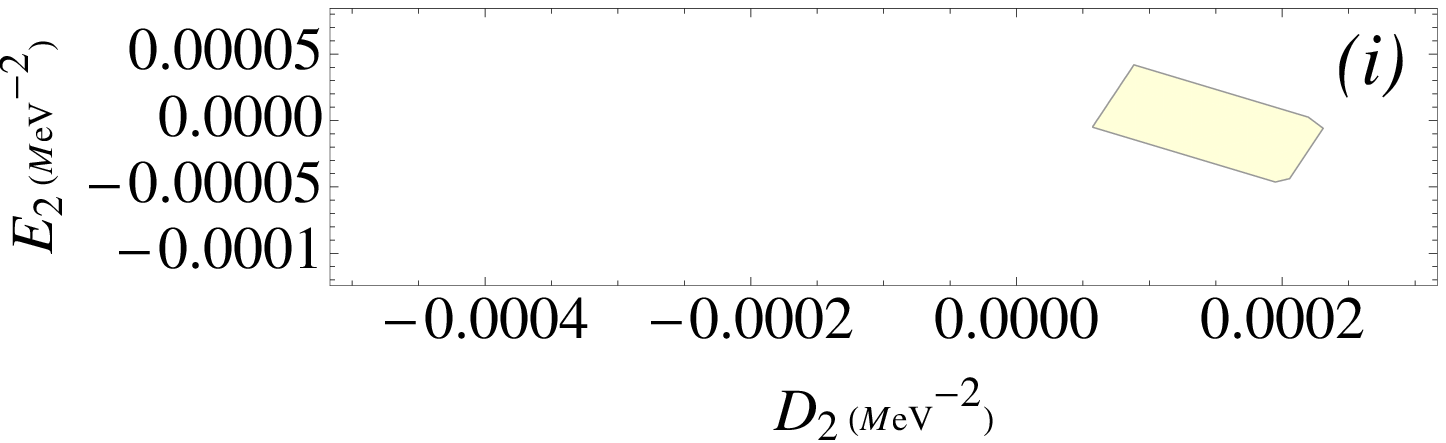}
	\includegraphics[width=0.8\textwidth]{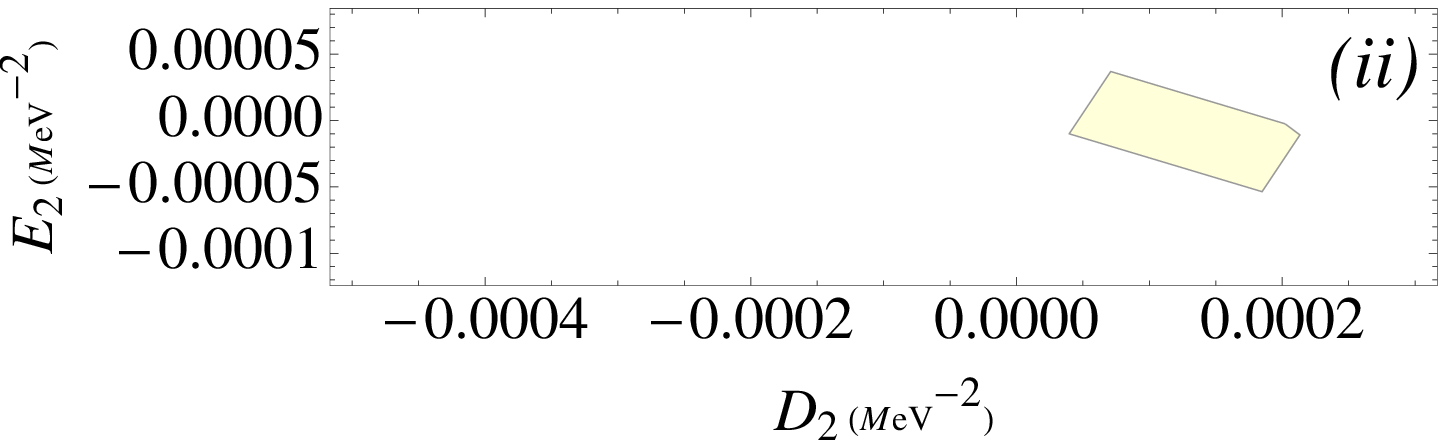} \\
	\includegraphics[width=0.8\textwidth]{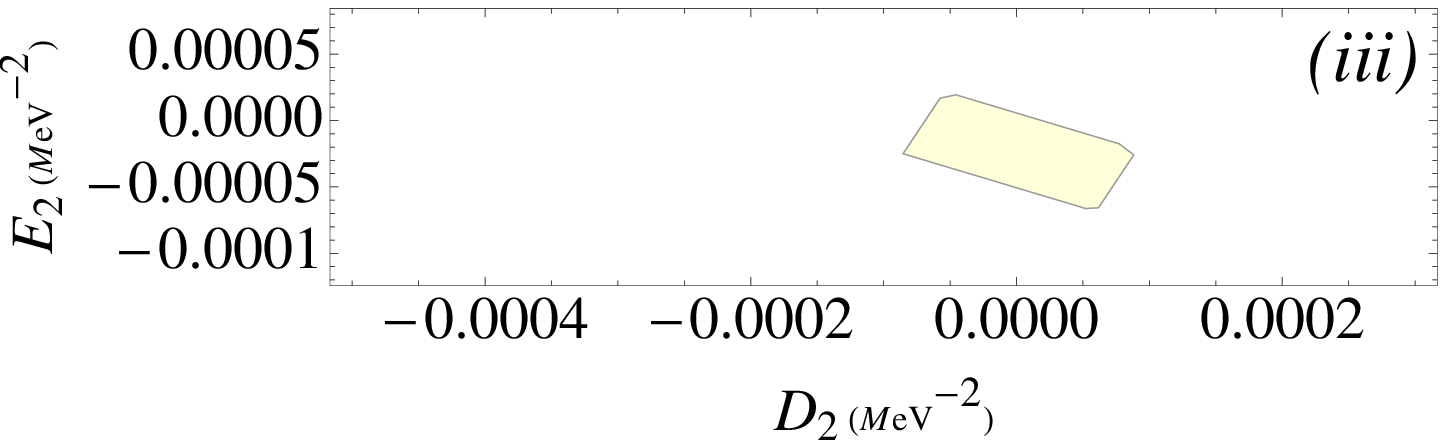}
	\includegraphics[width=0.8\textwidth]{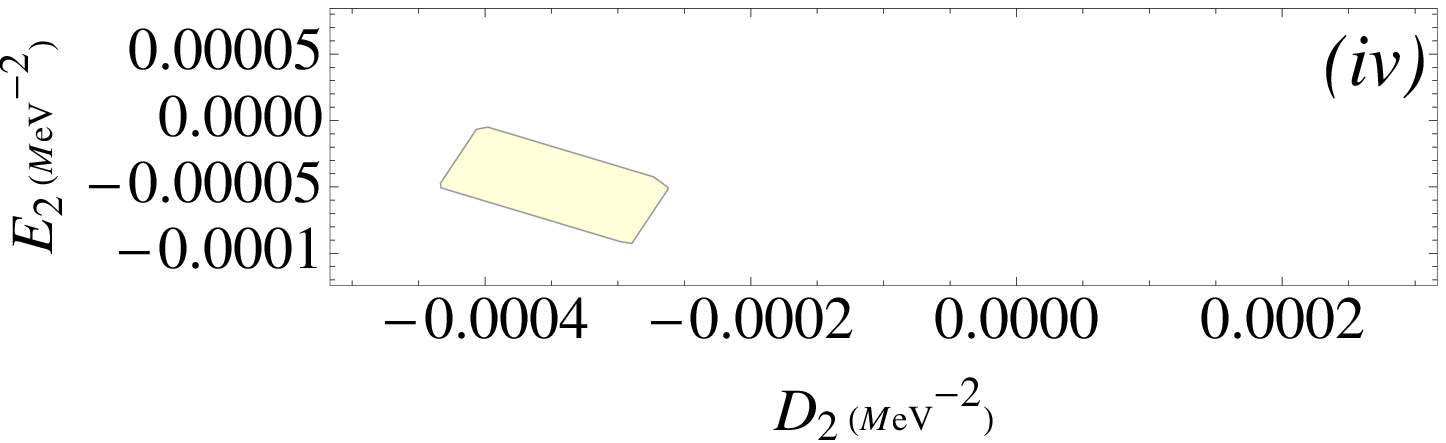}
	\caption{$(D_2,E_2)$-parameter 
space for the $D^{*} B^{*}({}^1 S_0)_{\xi}$ system ($\xi = s1, s2, t, d1, d2$).
Light shaded areas represent the regions in which the parameters
 acquire values that allow bound states 
for the nine states of $SU(3)_V$ light-quark flavor 
basis shown in Eq. (\ref{C2}), with binding energy greater 
than 0.1 MeV and obeying the condition $a_S \gtrsim 3 \lambda _{\pi}$. 
The legends in the  
panels represent respectively the 
choices $(i), (ii), (iii)$ and $(iv)$ done in previous Section for the  parameters 
$(D_1,E_1)$. }
	\label{FIGY0}
\end{figure}

\begin{figure}[htbp] 
	\centering
	\centering
	\includegraphics[width=0.8\textwidth]{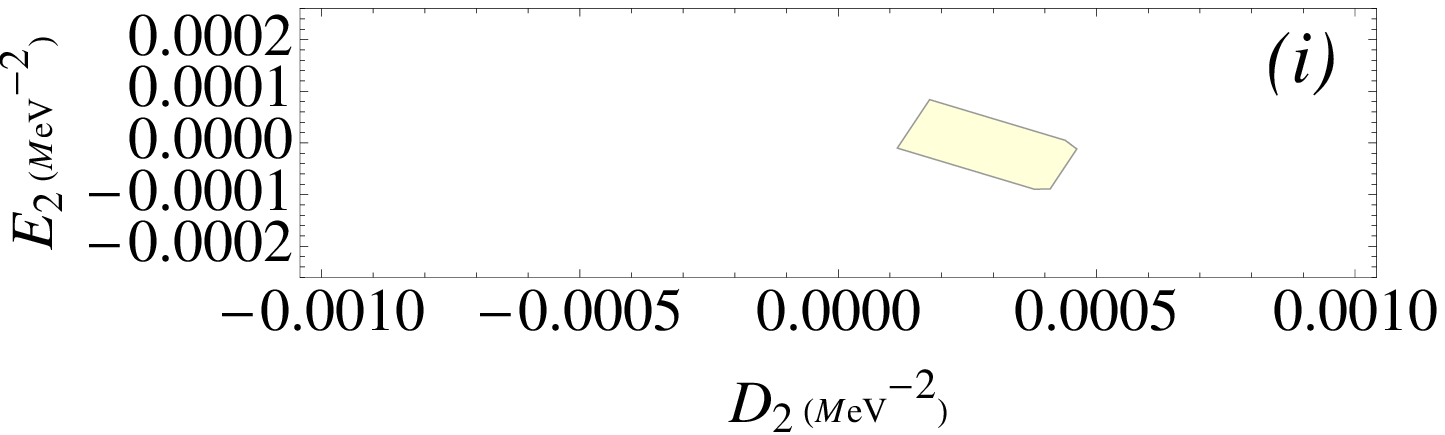}
	\includegraphics[width=0.8\textwidth]{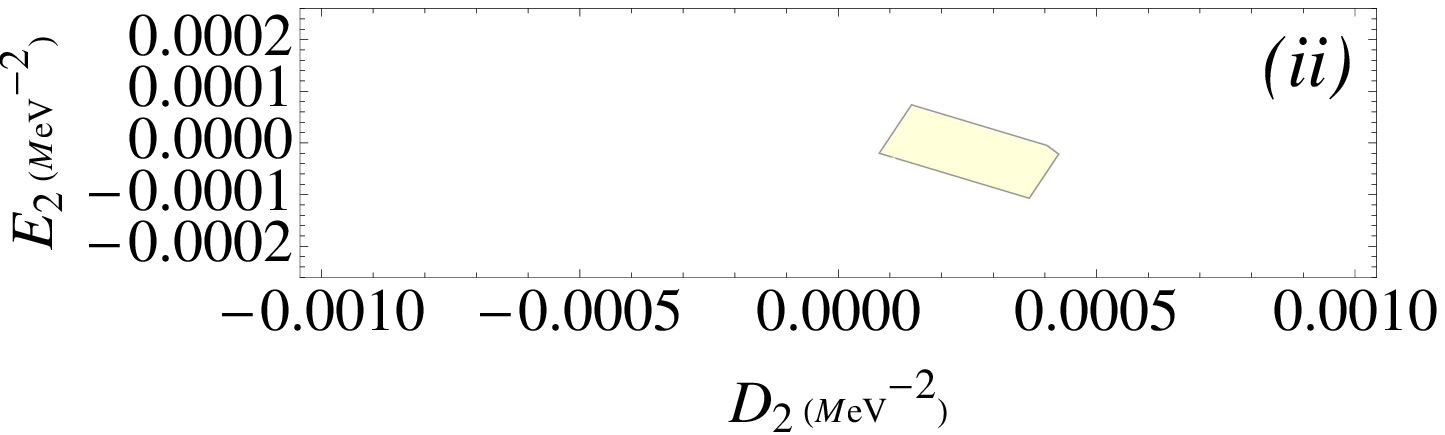} \\
	\includegraphics[width=0.8\textwidth]{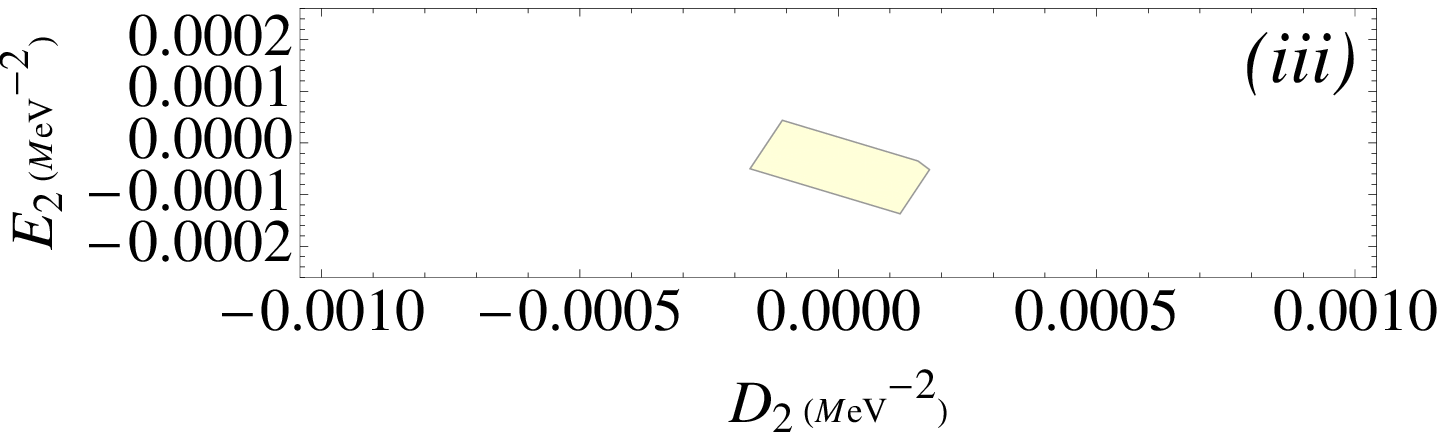}
	\includegraphics[width=0.8\textwidth]{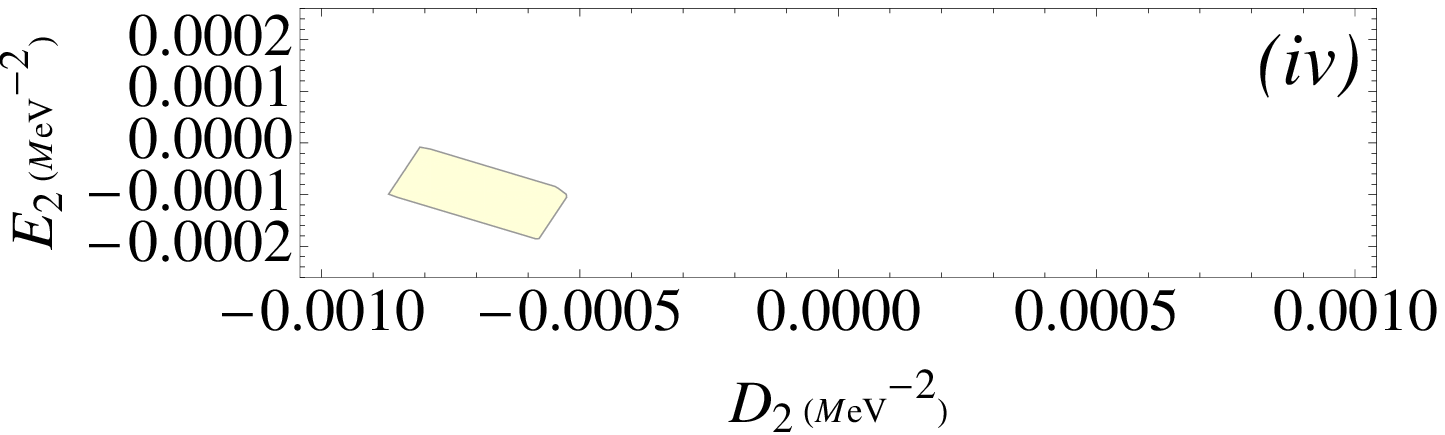}
	\caption{ Same as in Fig. \ref{FIGY0} 
	for $D^{*} B^{*}({}^3 S_1)_{\xi}$ system. }
	\label{FIGY1}
\end{figure}

\begin{figure}[htbp] 
	\centering
	\includegraphics[width=0.8\textwidth]{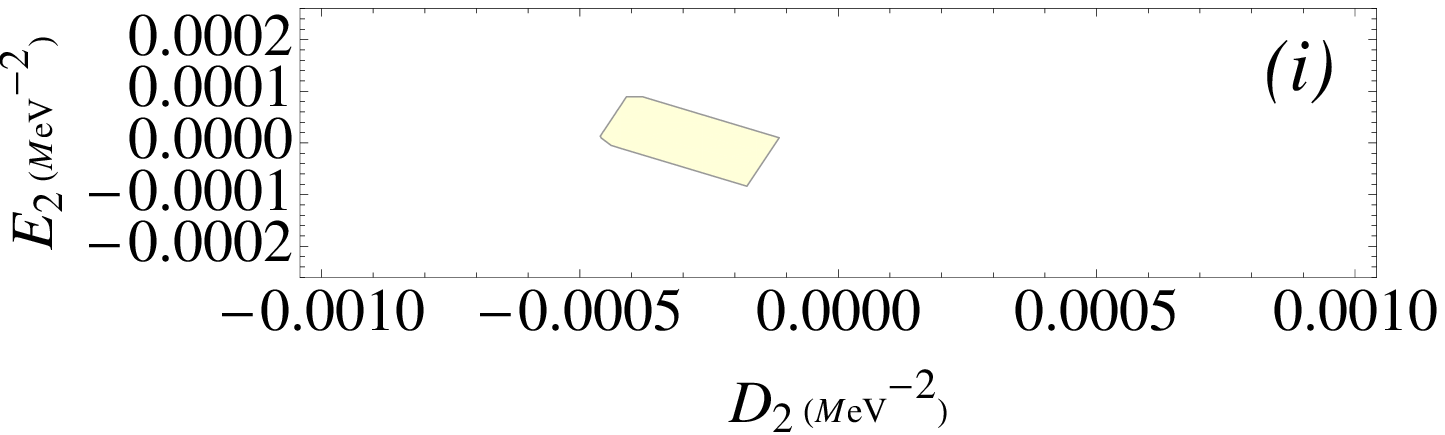}
	\includegraphics[width=0.8\textwidth]{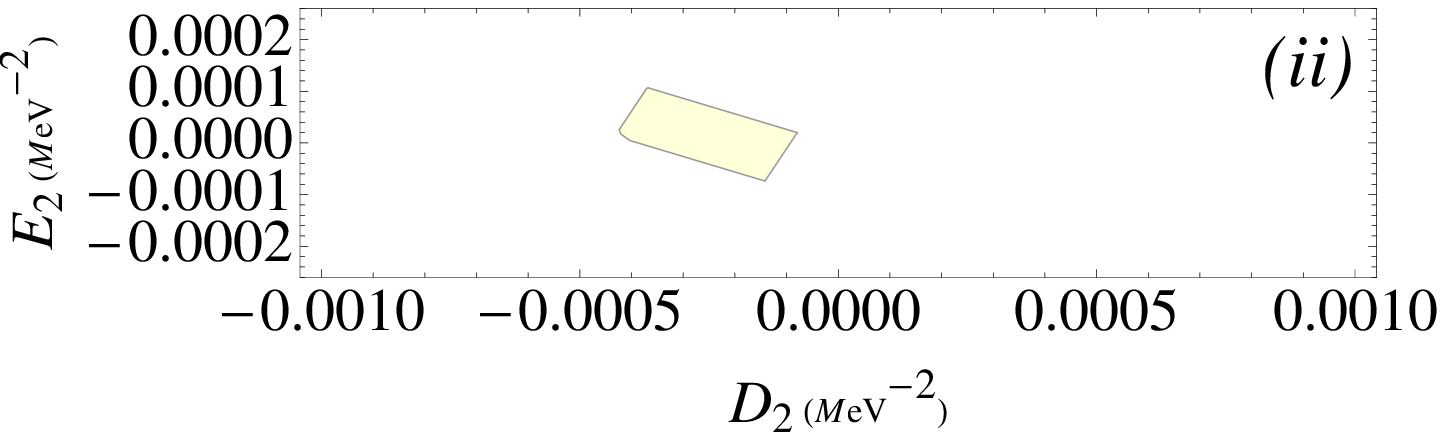} \\
	\includegraphics[width=0.8\textwidth]{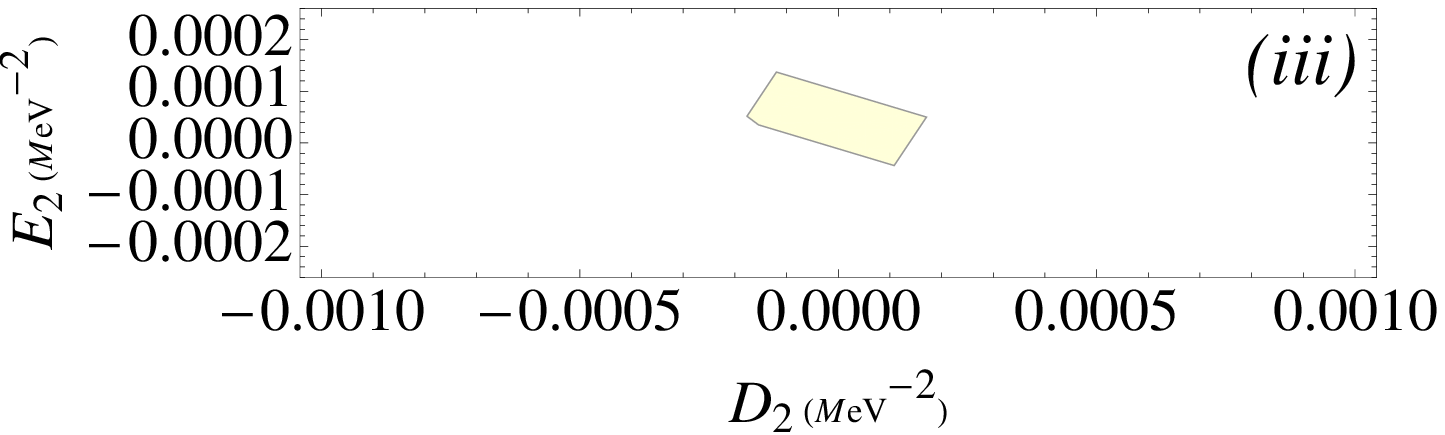}
	\includegraphics[width=0.8\textwidth]{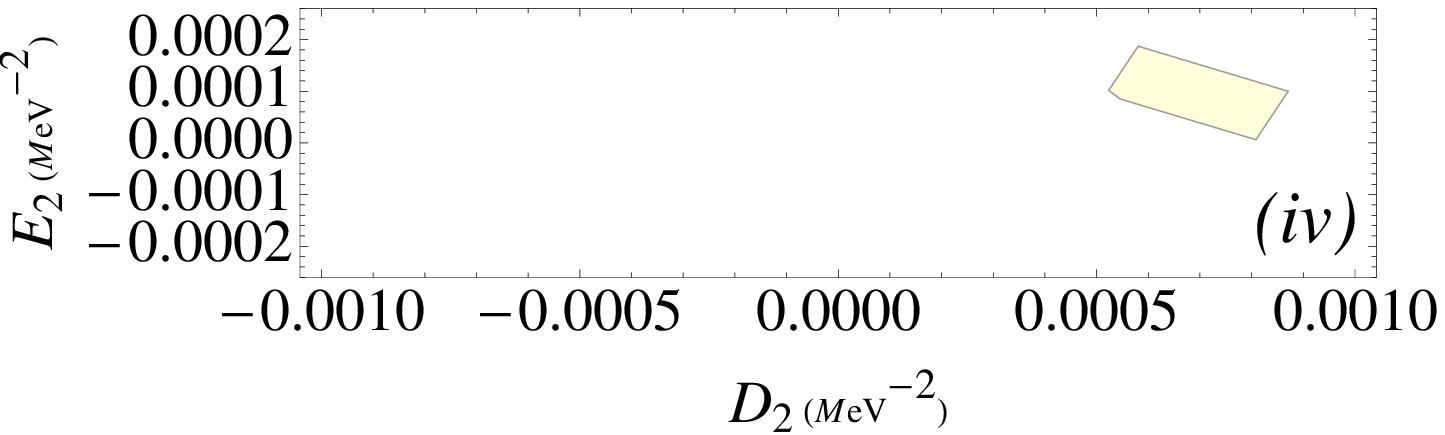}
	\caption{ Same as in Fig. \ref{FIGY0} 
	for $D^{*} B^{*}({}^5 S_2)_{\xi}$ system. }
	\label{FIGY2}
\end{figure}

First, we analyze the dependence of bound state solutions with the parameters 
$(D_2,E_2)$. In Figs. \ref{FIGY0}, \ref{FIGY1} and 
 \ref{FIGY2} are displayed the $(D_2,E_2)$-parameter 
space for the $D^{*} B^{*}({}^1 S_0)_{\xi}$, $D^{*} B^{*}({}^3 S_1)_{\xi}$ and 
  $D^{*} B^{*}({}^5 S_2)_{\xi}$ systems ($\xi = s1, s2, t, d1, d2$), respectively, 
 taking into account the different choices $(i)-(iv)$ of parameters 
$(D_1,E_1)$ done in previous Section. Light shaded areas represent the regions 
in which the parameters
 acquire values that allow bound states for the nine states of $SU(3)_V$ 
 light-quark flavor basis shown in Eq. (\ref{C2}), with binding energy greater 
than 0.1 MeV and obeying the condition $a_S \gtrsim 3 \lambda _{\pi}$.

These Figures indicate that the change of  $(D_1,E_1)$ parameters performs a 
displacement of the light-shaded areas, but without modification of their shapes 
and surfaces. We remark that in the case 
of $D^{*} B^{*}({}^1 S_0)_{\xi}$, $D^{*} B^{*}({}^3 S_1)_{\xi}$ systems, 
the increasing of magnitude of $(D_1,E_1)$ induces 
a decreasing of values of $(D_2,E_2)$ parameters to get bound states. On the other 
hand, the  $D^{*} B^{*}({}^5 S_2)_{\xi}$ system presents an inverse behavior: greater
values of  $(D_2,E_2)$ are necessary to yield bound-state solutions as the magnitudes 
$(D_1,E_1)$ grow.

\begin{figure}[htbp] 
	\centering
	\includegraphics[width=0.8\textwidth]{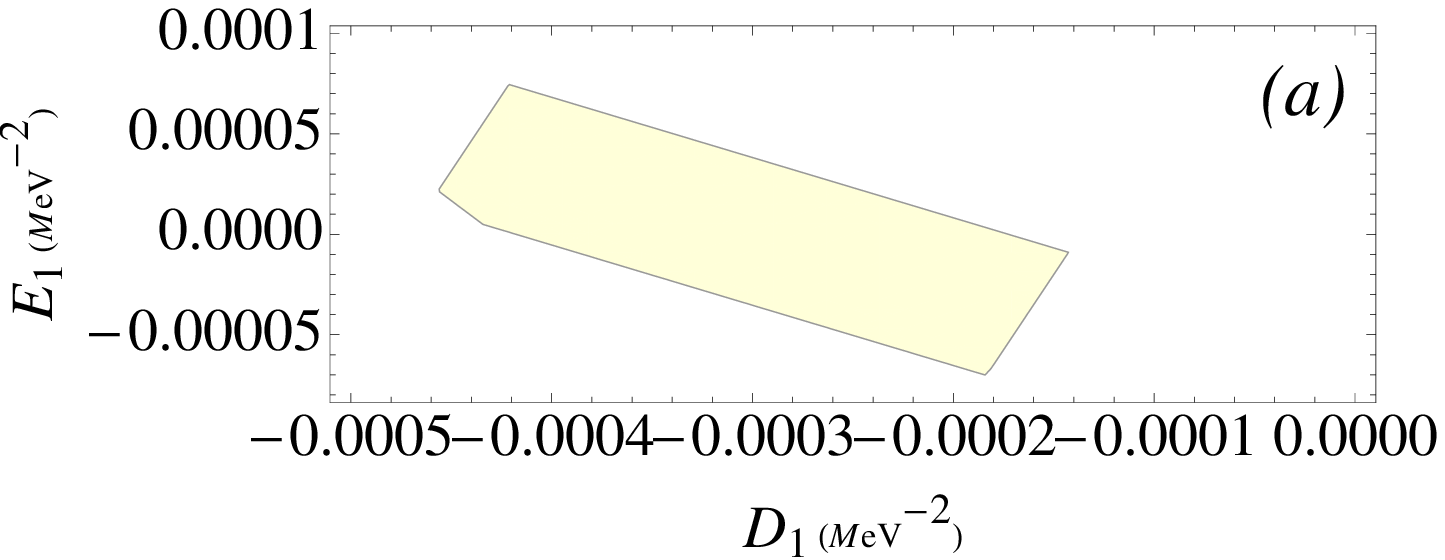}\\
	\includegraphics[width=0.8\textwidth]{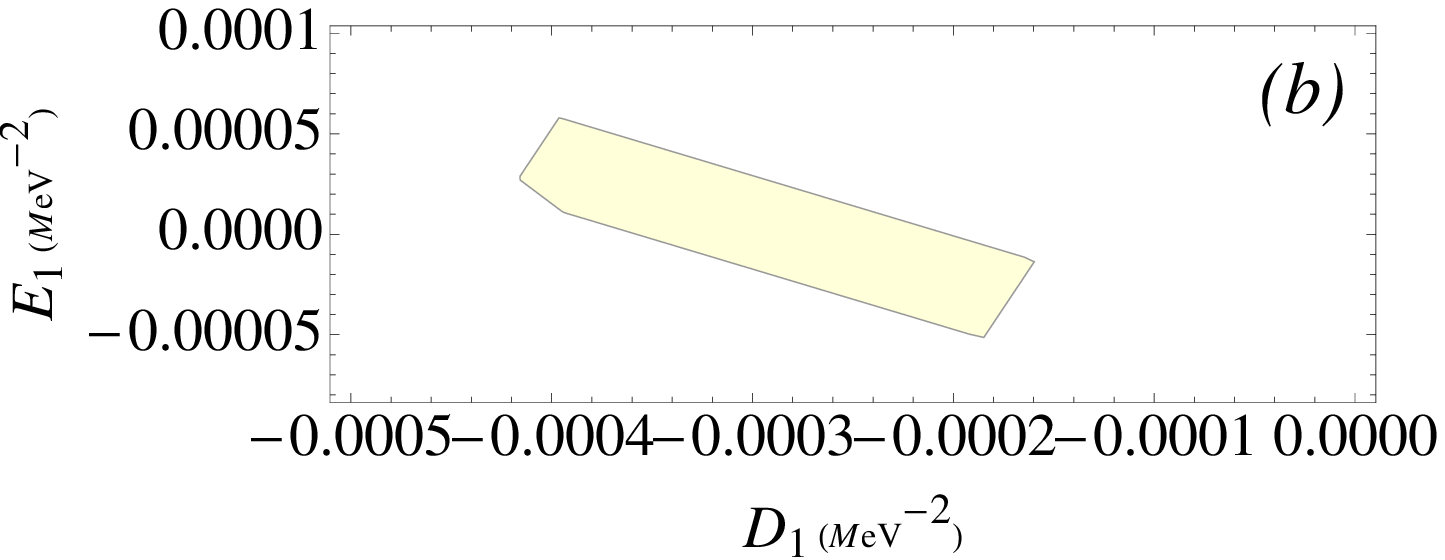} \\
	\centering
	\includegraphics[width=0.8\textwidth]{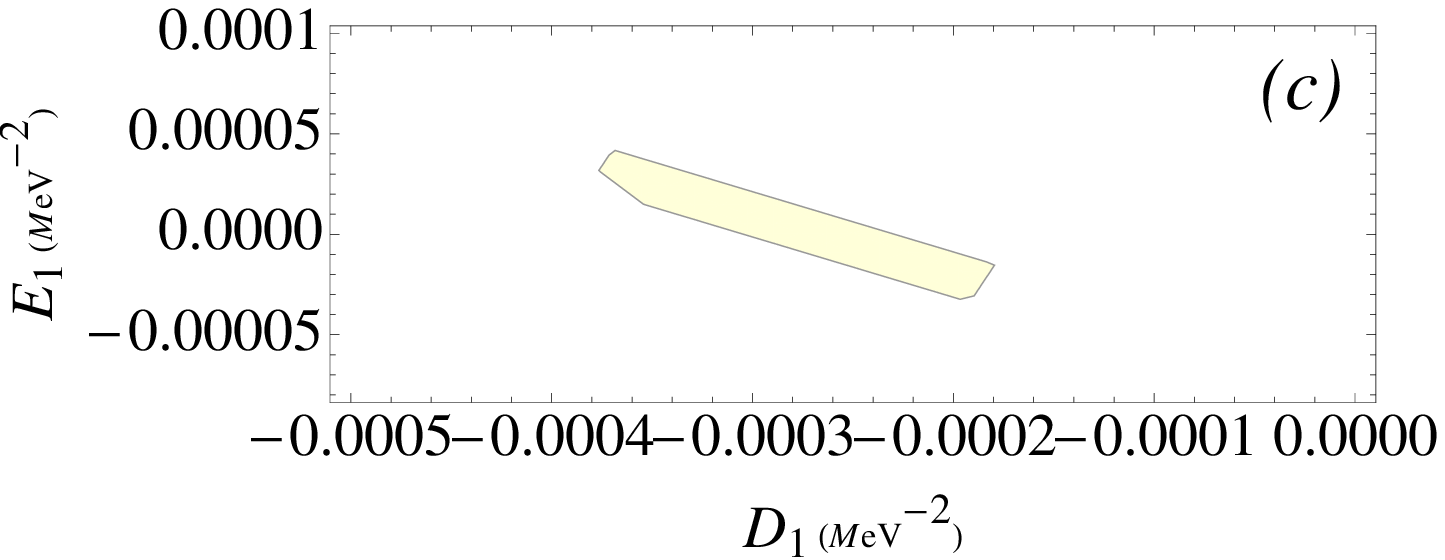}
\caption{$(D_1,E_1)$-parameter 
space. The light shaded areas indicate the 
intersection region in which bound states are obtained for the six studied systems 
($D B({}^1 S_0)_{\xi}$,
  $D^{*} B({}^3 S_1)_{\xi}$, $D B^{*}({}^3 S_1)_{\xi}$,
  $D^{*} B^{*}({}^1 S_0)_{\xi}$, $D^{*} B^{*}({}^3 S_1)_{\xi}$ and 
  $D^{*} B^{*}({}^5 S_2)_{\xi}$), , with binding energy greater 
than 0.1 MeV and obeying the condition $a_S \gtrsim 3 \lambda _{\pi}$. 
$\xi$ represents the nine flavor-state 
basis shown in Eq. (\ref{C2}). The legends in the  
panels represent respectively the 
choices $(a), (b)$ and $(c)$ for the  parameters 
$(D_2,E_2)$: 
$(a) D_2 = 0.00001\; \mathrm{MeV}^{-2}, \;
E_2= 0.00001\; \mathrm{MeV}^{-2};$ $(b) D_2 = 0.00005\; \mathrm{MeV}^{-2},
\; E_2= 0.000015\; \mathrm{MeV}^{-2};$
$(c) D_2 = 0.00008\; \mathrm{MeV}^{-2},\; E_2= 0.00002\; \mathrm{MeV}^{-2}$. }
	\label{FIGXZZZY0Y1Y2}
\end{figure}

Pursuing our analysis, In Fig. \ref{FIGXZZZY0Y1Y2} it is shown the 
$(D_1,E_1)$-parameter space for the following values of $(D_2,E_2)$:
\begin{description}
\item[$(a)$] $D_2 = 0.00001 \; \mathrm{MeV}^{-2}, \;E_2= 0.00001
\; \mathrm{MeV}^{-2};$
\item[$(b)$] $D_2 = 0.00005\; \mathrm{MeV}^{-2},\; E_2= 0.000015
\; \mathrm{MeV}^{-2};$ 
\item[$(c)$]  $D_2 = 0.00008\; \mathrm{MeV}^{-2}, E_2= 0.00002\; \mathrm{MeV}^{-2}.$
\end{description}
The light shaded areas in Fig. \ref{FIGXZZZY0Y1Y2} indicate the 
intersection region in which bound states are obtained
for the six studied systems ($D B({}^1 S_0)_{\xi}$,
  $D^{*} B({}^3 S_1)_{\xi}$, $D B^{*}({}^3 S_1)_{\xi}$,
  $D^{*} B^{*}({}^1 S_0)_{\xi}$, $D^{*} B^{*}({}^3 S_1)_{\xi}$ and 
  $D^{*} B^{*}({}^5 S_2)_{\xi}$), with binding energy greater 
than 0.1 MeV and obeying the condition $a_S \gtrsim 3 \lambda _{\pi}$.
The plots suggest that bound-state solutions are inhibited as 
the magnitude of the constants $(D_2,E_2)$ increases.

Taking into account the discussion above, we study the mass, binding energy and 
scattering length of the $D^{*} B^{*}$ systems for $(i)-(iv)$ choices of the 
parameters $(D_1,E_1)$ as in previous section, and at fixed values of $D2=0.00001
 \mathrm{MeV}^{-2} ; 
E2=0.00002  \mathrm{MeV}^{-2}$, chosen in order 
to reproduce the situation of upper-left panel in Fig. \ref{FIGY1}.    
The results are shown in Tables \ref{table5}, \ref{table6} and \ref{table7} in 
Appendix \ref{AppB}. We find loosely-bound state 
solutions in the following cases: 

\begin{enumerate}

\item $D^{*} B^{*}({}^1 S_0)$: there are loosely bound-state solutions for the 
 $SU(3)_V$ light-quark flavor isosinglets $s1$ and $s2$ in the situations
  $(i)$, $(ii)$ and $(ii)$; for the states $\xi =  t, d1, d2$ in the case $(iii)$.

\item $ D^{*} B^{*}({}^3 S_1)$: there are loosely bound-state solutions for the 
 $SU(3)_V$ light-quark flavor isosinglets $s1$ and $s2$ in the situation
 $(ii)$ and $(iii)$; for the isosinglet $s1$ in the case $(i)$; and for 
 the states $\xi = t, d1, d2$ in the case $(iii)$.

\item $D^{*} B^{*}({}^5 S_2)$: bound-state solution exists for all nine 
$ SU(3)$ flavor 
states ($\xi = s1, s2, t, d1, d2$) only in the situation $(iii)$.
\end{enumerate}
  
Also, as in the previous Section, the results suggest that for $D^{*} B^{*}({}^1 S_0)$ and 
$ D^{*} B^{*}({}^3 S_1)$ systems, 
the $t,d1,d2$ channels present greater binding energies than $s1, s2$  
in a specific choice of parameters. In the case of $D^{*} B^{*}({}^5 S_2)$, 
the binding energies are of the same order for a specific set of values of the 
parameters. Finally, it is important to notice that the number of bound-state 
solutions shown in Tables \ref{table5}-\ref{table7} could be lowered for greater 
values of the parameters $(D_2,E_2)$, as indicated in Fig. \ref{FIGXZZZY0Y1Y2}.

\subsection{Estimation of Interactions From Experimental Data
in Charmonium and Bottomonium Sectors}

Although there is no experimental information of exotic states in $B_c$ sector, 
it is also interesting at this point to analyze the parameter space in the light 
of available experimental data in charmonium and bottomonium sectors.  
Accordingly, in consonance with Refs. \cite{Hidalgo,Guo} we can use the 
information available of the $X(3872)$ and $Z_b (10610)$ masses as inputs 
to fix two of the four couplings. 

In this sense, we use the analogous approach discussed 
in previous Sections, by interpreting $X(3872)$  as $S$-wave isoscalar ($J^{PC} = 
 1^{++}$)  $D^0 \bar{D}^{* 0 }$ molecule (the charged conjugated particles 
 are implicit included) \cite{Esposito}. This 
 channel has 
 the leading-order potential given by $V^{(D^0 \bar{D}^{* 0 })}({}^3 S_1) 
 = C_1 + C_2$, with $C_1 = D_1 + 3  E_1, \,C_2 = D_2 + 3 E_2$. In the calculations 
 we consider the following central values of $X(3872)$ mass, threshold and binding 
 energy: $M_X = 3871.69$ MeV  \cite{PDG}, $m_{D^0} + m_{\bar{D}^{* 0}} 
 = 3871.8$ MeV  \cite{PDG}, and $E_b = 0.11$ MeV \cite{Esposito}, respectively.

 Similarly, the  $Z_b (10610)$ is understood as $S$-wave ($J^{PC} = 
 1^{+ -}$) isovector $(B \bar{B}^{* })$ molecule. In this case, the leading-order 
 potential is given by $V^{(B \bar{B}^{* })}({}^3 S_1) 
 = C_1 - C_2$, with $C_1 = D_1 - E_1, \,C_2 = D_2 - E_2$. In alignment with 
 Ref. \cite{Guo}, the binding energy of $Z_b (10610)$ is assumed to be 
   $2.0\pm 2.0$ MeV, while its mass and threshold are $M_{Z_b} = 10602.6\pm 2.0$ 
   MeV and  $m_{B} + m_{\bar{B}^{*} }
 = 10604.6$ MeV  \cite{PDG}, respectively. 

Therefore, using the informations of $X(3872)$ and $Z_b 
(10610)$ in Eqs. (\ref{BE1}) and (\ref{SL}), we obtain the following relations
\ben
D_1 = -0.00051477 + 0.5 D_2 - 1.5 E_2, \nonumber \\
E_1 =  -0.000422711 - 0.5 D_2 -   0.5 E_2. 
  \label{DE}  
 \een
  
To estimate the two remaining counterterms, we need two more states with 
different dependence of the couplings with respect to $X(3872)$ and $Z_b 
(10610)$. Following Ref. \cite{Hidalgo}, we consider the $X (3915)$ as a $ 0^{++}$
isoscalar $(D^{* } \bar{D}^{* })$ molecule, and $Y (4140)$ as a $ 0^{++}$
 $(D_s ^{* } \bar{D}_s ^{* })$ molecule, with the leading order potentials 
 given by $V^{(D^{* } \bar{D}^{* })}({}^1 S_0)  = (D_1 + 3 E_1) - 2 (D_2 + 3 E_2)$ and  
$V^{(D_s^{* } \bar{D}_s ^{* })}({}^1 S_0)  = (D_1 +  E_1) - 2 (D_2 +  E_2)$. The 
masses and thresholds are $M_{X(3915)} = 3917 $ MeV,
    $m_{\bar{D}^{*} } + m_{\bar{D}^{*} }
 = 4017.2$ MeV, $M_{Y(4140)} = 4140 $ MeV and
    $m_{\bar{D}^{*} } + m_{\bar{D}^{*} }
 = 4224.6$ MeV. Consequently, the use of these data in Eqs. (\ref{BE1}) and (\ref{SL}), 
 in combination with Eq. (\ref{DE}), yields 
 \ben 
 D_1=-0.0003973 \; \mathrm{MeV}^{-2}, \;\;  E_1 = 0.00027 
 \; \mathrm{MeV}^{-2}, \; \nonumber \\
 D_2 = -0.00017037\; \mathrm{MeV}^{-2}, \; E_2= 
 -0.0001351\; \mathrm{MeV}^{-2}. 
\label{DE2}
\een
Hence, with this specific renormalization scheme we can estimate the quantities of 
the $D^{(*)} B^{(*)}$ states. The results are displayed in 
Appendix \ref{AppC}. We remark that in this situation we have computed the total 
error by adding the partial errors (due to violations of heavy-quark spin and 
light flavor symmetries and in $Z_b(10610)$ binding energy)
in quadratures. We see in general a pattern of loosely bound states from these 
outcomes.

 It is 
worth remarking that the choice of $X (3915)$ and  $Y (4140)$ have some troubles, 
as greater binding energies and experimental status, as observed in Ref. 
\cite{Hidalgo}. In this sense, the predictions generated from this choice 
 will remain matter of debate. In addition, we stress that the 
renormalization procedure should be 
performed in each sector, 
and that there is not yet experimental evidence of hadronic molecules with both 
open charm and open bottom to have more solid predictions.

\section{Discussion and Concluding Remarks}
\lb{Conclusions}

In summary, we have investigated the interaction between charmed and bottomed mesons  
using Heavy-Meson Effective Theory. In this scenario we have discussed the 
formation deuteron-like molecules with 
both open charm and bottom. 
We have explored the regions in parameter space of the coupling constants in which the 
formation of bound $D^{(*)}  B^{(*)} $-states are allowed. 
Estimations of their masses, binding energies and scattering lengths have been 
performed as functions of interaction strength in a specific renormalization 
scheme. We  have
worked here in a specific renormalization scheme, in which the relation between 
the bare coupling constants is conserved after renormalization. This approach 
has allowed to relate the results for different sectors, taking into account the fact 
that there is not yet experimental evidence of hadronic molecules with both open 
charm and open bottom.

The analysis of bound-state solutions of $D^{(*)} B^{(*)}$ has been restricted 
to the region of 
relevance of contact-range interaction, in which the pion-exchange 
contribution is not relevant. In this sense, we have studied bound states which obey 
the condition $a_S \gtrsim 3 \lambda _{\pi}$.

The study is simpler and more direct for $DB({}^1 S_0)  ,
  D^{*}B({}^3 S_1)  $ and $  D B^{*} ({}^3 S_1) $ systems, since 
they depend on two of the four coupling constants in Effective Lagrangian. 
In this context, the analysis of the parameter space has indicated a region 
in which the relevant 
parameters acquire values that allow loosely bound states 
for some of the nine $ SU(3)_V$ light-quark flavor states
 (two isosinglets, one triplet and two doublets).
The results suggested that for both spin-0 and spin-1 systems, 
the $t,d1,d2$ channels present greater binding energies than $s1, s2$  
in a specific choice of parameters. 

For the $D^{*} B^{*}({}^1 S_0), D^{*} B^{*}({}^3 S_1)$ and 
$D^{*} B^{*}({}^5 S_2)$ systems, the parameter configuration space is richer, 
since the transition amplitudes depend on the four coupling constants. This study has
suggested, as in the above mentioned situation, that the existence of loosely bound 
states for the $ SU(3)_V$ light-quark flavor states
 strongly depends on the magnitude of the 
parameters. In the case 
of $D^{*} B^{*}({}^1 S_0)_{\xi}$, $D^{*} B^{*}({}^3 S_1)_{\xi}$ systems, for a 
specific choice of renormalization scheme 
the increasing of magnitude of $(D_1,E_1)$ induces 
a decreasing of values of $(D_2,E_2)$ parameters in order to get bound states; while 
 with
 $D^{*} B^{*}({}^5 S_2)_{\xi}$ system this dependence is inverse: greater
values of  $(D_2,E_2)$ are necessary to yield bound-state solutions.

In addition, the $D^{(*)} B^{(*)}$ states have also analyzed with a specific 
renormalization scheme, in which the couplings have been fixed by considering 
the location of $X(3872)$, $Z_b 
(10610)$, $X (3915)$ and $Y (4140)$ states, interpreted here as heavy-meson 
molecules. The results obtained with this specific choice reveal in general a pattern 
of loosely bound states.

We notice that some results reported above can be compared with other ones available
in literature. For example, the state $ | D^{*} B^{*} ({}^5 S_2)_{s1} \rangle$ 
with the parameters assuming values between the choices $(ii)$ and 
$(iii)$ displayed in table \ref{table7} has its mass
in the range of the outcome obtained in Ref. \cite{Guo}. In addition, the mass 
yielded for the state $ | D^{*} B^{*} ({}^3 S_1)_{t} \rangle$ in the case 
$(iii)$
of Table \ref{table6}, is in the range of the equivalent state discussed in 
Ref. \cite{Guo}. If we compare with the 
specific renormalization scheme shown in Eq. (\ref{DE2}), fixed to location of 
observed 
exotic states in charmonium and bottomonium sectors, we see that our findings 
in Table \ref{table8} 
for the $| D^{*} B^{*} ({}^3 S_1)_{t} \rangle$ state  are in the range of the 
results in Ref. \cite{Guo}, but in the case of $ | D^{*} B^{*} ({}^5 S_2)_{s1} \rangle$ 
our estimation yields a very loosely bound state.

Furthermore, bound-state solutions pointed out in Ref. \cite{Sun} 
 can be discussed in some sense under our point of 
view, with the parameters in the range of the choices $(i)$ and $(ii)$ of 
Tables \ref{table2}-\ref{table7}. Notice, however, that  Ref. \cite{Sun} 
has considered mixing S-D mixing effect, and we must be careful in this comparison.

Further work is needed to improve these results, in order to perform more precise 
comparison with other phenomenological models, and to contribute to the 
experimental search of $D^{(*)}  B^{(*)} $ molecular states. It is possible, for
example, to study the decays of these predicted molecular states into $B_c$ mesons 
and light mesons, as suggested in Ref. \cite{Sun}. Another improvement is the 
inclusion of the pion-exchange potential, in order to extend the range of 
applicability of this approach.

\begin{center}
{\bf ACKNOWLEDGMENTS}
\end{center}

We thank CAPES and CNPq  (Brazilian Agencies) for financial support.

\appendix
\section[Appendix]{Appendix}

\lb{Appendix}

\subsection{Quantities for the $DB({}^1 S_0)_{\xi}, D^{*}
B({}^3 S_1)_{\xi}$ and $D B^{*} ({}^3 S_1)_{\xi}$ systems for different values of 
$(D_1, E_1)$}
\label{AppA}

\begin{longtable}{cccccc}
 \caption{ Relevant quantities for the $DB({}^1 S_0)_{\xi}$ system
 ($\xi = s1, s2, t, d1, d2$ represent the nine flavor states shown in Eq.
  (\ref{C2})). $M_{Th}$, $M$, B.E., and $a_s$ mean threshold, mass, 
 binding energy and 
scattering length of the respective state. The parameters $(D_1, E_1)$ are chosen
in the four situations: $(i) D_1=-0.000015 \; \mathrm{MeV}^{-2}, \;
E_1=-0.00001\; \mathrm{MeV}^{-2}; (ii) D_1=-0.00005\; \mathrm{MeV}^{-2},
\; E_1=-0.00002\; \mathrm{MeV}^{-2}; 
(iii)  D_1=-0.0003\; \mathrm{MeV}^{-2}, E_1=-0.00005\; \mathrm{MeV}^{-2};
(iv)  D_1=-0.001\; \mathrm{MeV}^{-2}, \;E_1=-0.0001\; \mathrm{MeV}^{-2}. $
 Bold columns indicate states with binding energy greater 
than 0.1 MeV and obeying the condition $a_S \gtrsim 3 \lambda _{\pi}$.} \\
\hline
 $\xi$ & $M_{Th}$ (MeV) & $(D_1, E_1)$ & $M$ (MeV) & B.E. (MeV) & $a_s$ (fm) \\
\hline
$s1 $ & \multirow{4}{*}{7146.6} 
 & (i) & 7133.7 $\pm$ 3.9 & 12.9 $\pm $3.9 & 1.1 $\pm$ 0.2  \\
&  & \textbf{(ii)}    &  \textbf{7144.4 $\pm$ 0.7} & \textbf{2.2 $\pm$ 0.7} & 
\textbf{2.6 $\pm$ 0.4}   \\
&  & \textbf{(iii)} &  \textbf{7146.5 $\pm$ 0.1 }  & \textbf{0.1 $\pm$0.1} 
& \textbf{10.3 $\pm$ 1.5}  \\
& & (iv) & 7146.58 $\pm$ 0.01 & 0.02 $\pm$ 0.01 & 29.3 $\pm$ 4.4  \\
 \hline 
 ${s2}$ & \multirow{4}{*}{ 7335.3 } 
 & (i) &  7302.4 $\pm$ 16.5 &  32.9$\pm$ 16.5  & 0.7$\pm$ 0.2 \\
 &    & (ii) & 7330.8 $\pm$ 2.3 & 4.5 $\pm$ 2.3 & 1.8 $\pm$ 0.4\\
 &    & (iii) & \textbf{7335.1$\pm$ 0.1} &  \textbf{0.2 $\pm$ 0.1} 
 & \textbf{8.4 $\pm$ 2.1}  \\
&  & (iv)   & 7335.28 $\pm$ 0.01 &  0.02 $\pm$ 0.01 & 26 $\pm$ 6.5  \\
 \hline 
 ${t}$ & \multirow{4}{*}{ 7146.6 } 
  & (i) & 6713.4 $\pm$ 130  &  433.2 $\pm$ 130 & 0.2$\pm$ 0.02  \\
&   & (ii) & 7124.2 $\pm$ 6.7  &  22.4 $\pm$ 6.7 & 0.8$\pm$ 0.1 \\
&    & \textbf{(iii)} & \textbf{7146.2 $\pm$ 0.2}  & \textbf{0.4 $\pm$ 0.2} 
& \textbf{5.9 $\pm$ 0.8} \\
&  & (iv)  & 7146.57 $\pm$ 0.01  & 0.03 $\pm$ 0.01 & 20.5 $\pm$ 3 \\
 \hline  
 ${d1}$ & \multirow{4}{*}{ 7247.9 } 
  & (i)   & 6862.2 $\pm$ 192 & 385.7 $\pm$ 192 & 0.2 $\pm$ 0.1\\
&   & (ii)  & 7223 $\pm$ 10 & 19.9 $\pm$ 10 & 0.8 $\pm$ 0.2 \\
&   & \textbf{(iii)} & \textbf{7247.5} $\pm$ 0.2 & \textbf{0.4} $\pm$ 0.2 
& \textbf{6.1} $\pm$ 1.5   \\
&  & (iv)   & 7247.87 $\pm$ 0.01 & 0.03 $\pm$ 0.01 & 21.3 $\pm$ 5.3 \\
 \hline 
${d2}$ & \multirow{4}{*}{ 7234 } 
   & (i)   & 6806.3 $\pm$ 213 & 427.7 $\pm$ 213 & 0.2 $\pm$ 0.1 \\
&   & (ii)  & 7211.9 $\pm$ 11 & 22.1 $\pm$ 11 & 0.8  $\pm$ 0.2  \\
&     & \textbf{(iii)} & \textbf{7233.6  $\pm$ 0.2} & \textbf{0.4  $\pm$ 0.2}
 & \textbf{5.9  $\pm$ 1.5} \\
&   & (iv)  & 7233.97  $\pm$ 0.02 & 0.03  $\pm$ 0.2& 20.6  $\pm$ 5.1 \\
 \hline 
  \hline 
\label{table2}
 \end{longtable}
 
\begin{longtable}{cccccc}   
 \caption{ Same as in table \ref{table2} for $D^{*}B({}^3 S_1)_{\xi}$ system.} 
 \\
\hline
$\xi$ & $M_{Th}$ (MeV) & $(D_1, E_1)$ & $M$(MeV) & B.E. (MeV) & $a_s$ (fm) \\
\hline
${s1}$ & \multirow{4}{*}{7288.}
 & (i)  &  7277 $\pm$ 3.3 & 10.972 $\pm$ 11 & 1.1  $\pm$ 0.2 \\
&  & \textbf{(ii)} &  \textbf{7286.1 $\pm$ 0.6} & \textbf{1.9  $\pm$ 0.6}
 & \textbf{2.7  $\pm$ 0.4}  \\
& & \textbf{(iii)} &  \textbf{7287.9  $\pm$ 0.04} & \textbf{0.1}  $\pm$ 0.04
 & \textbf{10.8 $\pm$ 1.6}  \\
& & (iv)  &  7287.99  $\pm$ 0.01 & 0.01 $\pm$ 0.01 & 30.9  $\pm$ 4.6 \\
 \hline 
 ${s2}$ & \multirow{4}{*}{ 7479.1 } 
  & (i) &  7450.8  $\pm$ 14.1 & 28.2$\pm$ 14.1 & 0.7  $\pm$ 0.2 \\
&   & (ii) & 7475.2 $\pm$ 1.9 & 3.9 $\pm$ 1.9 & 1.5 $\pm$ 0.5 \\
&   & \textbf{(iii)} & \textbf{7478.9 $\pm$ 0.1} & \textbf{0.2  $\pm$ 0.1}
 & \textbf{8.8  $\pm$ 2.2}  \\
&  & (iv)   & 7479.09  $\pm$ 0.01& 0.01  $\pm$ 0.01 & 27.3 $\pm$ 6.8  \\
 \hline 
 ${t}$ & \multirow{4}{*}{ 7288 } 
  & (i) & 6918.9  $\pm$ 110.7 & 369.1 $\pm$ 110.7 & 0.2 $\pm$ 0.1 \\
&   & (ii) & 7268.9 $\pm$ 5.7 & 19.1 $\pm$ 5.7 & 0.9  $\pm$ 0.1\\
&   & \textbf{(iii)} & \textbf{7287.6 $\pm$ 0.1} & \textbf{0.4 $\pm$ 0.1}
 & \textbf{6.2 $\pm$ 0.9} \\
&  & (iv)  & 7287.97  $\pm$ 0.01& 0.03  $\pm$ 0.01 & 21.6  $\pm$ 3.2\\
 \hline  
 ${d1}$ & \multirow{4}{*}{ 7391.7 } 
  & (i) & 7060.6 $\pm$ 165.6 &  331.1  $\pm$ 165.6 & 0.2 $\pm$ 0.1 \\
&  & (ii)  & 7374.6 $\pm$ 8.6 & 17.1 $\pm$ 8.6  & 0.9 $\pm$ 0.2 \\
&   & \textbf{(iii)} & \textbf{7391.4 $\pm$ 0.2} & \textbf{0.3 $\pm$ 0.2} 
& \textbf{6.4 $\pm$ 1.6}  \\
&  & (iv)   & 7391.67 $\pm$ 0.01 & 0.03 $\pm$ 0.01 & 22.4 $\pm$ 5.6 \\
 \hline 
${d2}$ & \multirow{4}{*}{ 7375.4 } 
   & (i)   & 7011.3 $\pm$ 182.1 & 364.1 $\pm$ 182.1 & 0.2 $\pm$ 0.5 \\
&    & (ii)  & 7356.6 $\pm$ 9.4 & 18.8 $\pm$ 9.4 & 0.9 $\pm$ 0.2  \\
&    & \textbf{(iii)} & \textbf{7375 $\pm$ 0.2} & \textbf{0.4 $\pm$ 0.2} 
& \textbf{6.2 $\pm$ 2.6} \\
&   & (iv)  & 7375.37 $\pm$ 0.01 & 0.03 $\pm$ 0.1 & 21.7 $\pm$ 5.4 \\
 \hline 
   \hline 
\label{table3}
 \end{longtable}

\begin{longtable}{cccccc}
 \caption{ Same as in table \ref{table2} for $D B^{*}({}^3 S_1)_{\xi}$ system.} 
 \\
\hline
$\xi$ & $M_{Th}$ (MeV) & $(D_1, E_1)$ & $M$ (MeV) & B.E. (MeV) & $a_s$ (fm) \\
\hline
${s1}$ & \multirow{4}{*}{7192.4}
 & (i)  &  7179.6  $\pm$ 3.8 & 12.8 $\pm$ 3.8 & 1.1  $\pm$ 0.2 \\
&  & \textbf{(ii)} & \textbf{7190.2 $\pm$ 0.7} & \textbf{2.2 $\pm$ 0.7}
 & \textbf{2.6 $\pm$ 0.4}  \\
&  & \textbf{(iii)} &  \textbf{7192.3 $\pm$ 0.4 } & \textbf{0.1 $\pm$ 0.4}
 & \textbf{10.3 $\pm$ 1.5}  \\
& & (iv)  &  7192.38  $\pm$ 0.01 & 0.02 $\pm$ 0.1 & 29.3 $\pm$ 4.4 \\
 \hline 
 ${s2}$ & \multirow{4}{*}{ 7383.9 } 
 & (i) &  7351.2  $\pm$ 16.3 & 32.7  $\pm$ 16.3 & 0.7 $\pm$ 0.2\\
&  & (ii) & 7379.4  $\pm$ 2.2 & 4.5 $\pm$ 2.2 & 1.8 $\pm$ 0.4 \\
&   & \textbf{(iii)} & \textbf{7383.7 $\pm$ 0.1} & \textbf{0.2 $\pm$ 0.1} 
& \textbf{8.4 $\pm$ 2.1}  \\
&  & (iv)   & 7383.88 $\pm$ 0.01 & 0.02 $\pm$ 0.01 & 26 $\pm$ 6.5  \\
 \hline 
 ${t}$ & \multirow{4}{*}{ 7192.4 } 
  & (i) & 6762.1 $\pm$ 129.1 & 430.3  $\pm$ 129.1 & 0.2 $\pm$ 0.03 \\
&   & (ii) & 7170.2  $\pm$ 6.7 & 22.2 $\pm$ 6.7 & 0.8  $\pm$ 0.1\\
&   & \textbf{(iii)} & \textbf{7192 $\pm$ 0.1} & \textbf{0.4 $\pm$ 0.1} 
& \textbf{5.9 $\pm$ 0.9} \\
&  & (iv)  & 7192.37 $\pm$ 0.01 & 0.03 $\pm$ 0.01 & 20.5  $\pm$ 3.1\\
 \hline  
 ${d1}$ & \multirow{4}{*}{ 7293.7 } 
  & (i) & 6910.7  $\pm$ 191.5 & 383  $\pm$ 191.5 & 0.2 $\pm$ 0.1 \\
&  & (ii)  & 7273.9 $\pm$ 9.9 & 19.8 $\pm$ 9.9 & 0.8  $\pm$ 0.2\\
&   & \textbf{(iii)} & \textbf{7293.3 $\pm$ 0.2} & \textbf{0.4 $\pm$ 0.2} 
& \textbf{6.1 $\pm$ 1.5}  \\
&  & (iv)   & 7293.67 $\pm$ 0.2 & 0.03 $\pm$ 0.2 & 21.3 $\pm$ 5.3 \\
 \hline 
${d2}$ & \multirow{4}{*}{ 7282.6 } 
   & (i)   & 6857.8  $\pm$ 212.4 & 424.8 $\pm$ 212.4 & 0.2 $\pm$ 0.1 \\
&    & (ii)  & 7260.7  $\pm$ 11 & 21.9  $\pm$ 11 & 0.8  $\pm$ 0.2 \\
&    & \textbf{(iii)} & \textbf{7282.2 $\pm$ 0.2} & \textbf{0.4 $\pm$ 0.2}
 & \textbf{5.9 $\pm$ 1.5} \\
&   & (iv)  & 7282.57  $\pm$ 0.2 & 0.03  $\pm$ 0.2 & 20.6  $\pm$ 5.2 \\
 \hline 
   \hline 
\label{table4}
 \end{longtable}

\subsection{Quantities for the $D^{*} B^{*}({}^1 S_0)_{\xi}, D^{*} B^{*}({}^3 S_1)
_{\xi}$ and $D^{*} B^{*}({}^5 S_2)_{\xi}$ systems for different values of 
$(D_1, E_1)$}
 \label{AppB}

\begin{longtable}{cccccc}
 \caption{ Same as in table \ref{table2} for $D^{*} B^{*}({}^1 S_0)_{\xi}$ system, 
 with ($D2=0.00001\; \mathrm{MeV}^{-2} ;\; E2=0.00002\; \mathrm{MeV}^{-2}$).  } 
 \\
\hline
$\xi$ & $M_{Th}$ (MeV) & $(D_1, E_1)$ & $M$ (MeV) & B.E. (MeV) & $a_s$ (fm) \\
\hline
${s1}$ & \multirow{4}{*}{7333.8}
 & \textbf{(i)}  & \textbf{7333.2 $\pm$ 0.2}  &  \textbf{0.6 $\pm$ 0.2}
  &  \textbf{4.7 $\pm$ 0.7} \\
& & \textbf{(ii)} & \textbf{7333.5 $\pm$ 0.1} & \textbf{0.3 $\pm$ 0.1} 
&  \textbf{6.3 $\pm$ 0.9}  \\
& & \textbf{(iii)} & \textbf{7333.7  $\pm$ 0.2} & \textbf{ 0.1 $\pm$ 0.2}
 &  \textbf{14.4 $\pm$ 2.2 } \\
& & (iv)  &  7333.79  $\pm$ 0.01 &  0.01  $\pm$ 0.01 &  34.5 $\pm$ 5.2 \\
 \hline 
 ${s2}$ & \multirow{4}{*}{ 7527.7 } 
  & \textbf{(i)} &  \textbf{7525.5 $\pm$ 1.1} &  \textbf{2.2 $\pm$ 1.1} 
  &  \textbf{2.5 $\pm$ 0.6}\\
&   & \textbf{(ii)} & \textbf{7526.7 $\pm$ 0.5 } & \textbf{1 $\pm$ 0.5}
 &  \textbf{3.6 $\pm$ 0.9} \\
&  & \textbf{(iii)} & \textbf{7527.6 $\pm$ 0.1} &  \textbf{0.1 $\pm$ 0.1}
 &  \textbf{10.6 $\pm$ 2.7} \\
&  & (iv)   & 7527.68 $\pm$ 0.01 &  0.02  $\pm$ 0.01 &  29.2 $\pm$ 7.3 \\
 \hline 
 ${t}$ & \multirow{4}{*}{ 7333.8 } 
  & (i) & - &  - &  - \\
&  & (ii) & 7305.5 $\pm$ 8.5 &  28.3 $\pm$ 8.5 &  0.7 $\pm$ 0.1 \\
&   & \textbf{(iii)} & \textbf{7333.4 $\pm$ 0.1} &  \textbf{0.4 $\pm$ 0.1} 
&  \textbf{6 $\pm$ 0.9} \\
&  & (iv)  & 7333.77 $\pm$ 0.01 &  0.03 $\pm$ 0.01 &  21.5 $\pm$ 3.2 \\
 \hline  
 ${d1}$ & \multirow{4}{*}{ 7437.5 } 
  & (i) & - &  - &  - \\
&   & (ii)  & 7412.1 $\pm$ 12.7 &  25.4 $\pm$ 12.7 &  0.7 $\pm$ 0.2 \\
&   & \textbf{(iii)} & \textbf{7437.2 $\pm$ 0.2} &  \textbf{0.3 $\pm$ 0.2} 
&  \textbf{6.3 $\pm$ 1.6}  \\
&  & (iv)   & 7437.47 $\pm$ 0.01 &  0.03  $\pm$ 0.01 &  22.3  $\pm$ 5.6 \\
 \hline 
${d2}$ & \multirow{4}{*}{ 7424 } 
   & (i)   & - &  - & - \\
&   & (ii)  & 7396.1  $\pm$ 13.9 &  27.9  $\pm$ 13.9 &  0.7  $\pm$ 0.2 \\
&   & \textbf{(iii)} & \textbf{7423.6 $\pm$ 0.2} &  \textbf{0.4 $\pm$ 0.2} 
&  \textbf{6.1 $\pm$ 1.5} \\
&   & (iv)  & 7423.97 $\pm$ 0.01 &  0.03  $\pm$ 0.01 &  21.6 $\pm$ 5.4 \\
 \hline   \hline 
\label{table5}
 \end{longtable}

\begin{longtable}{cccccc}
 \caption{ Same as in table \ref{table2} for $D^{*} B^{*}({}^3 S_1)_{\xi}$ system, 
 with ($D2=0.00001 \; \mathrm{MeV}^{-2}; \;E2=0.00002\; \mathrm{MeV}^{-2}$).  } 
 \\
\hline
$\xi$ & $M_{Th}$ (MeV) & $(D_1, E_1)$ & $M$ (MeV) & B.E. (MeV) & $a_s$ (fm) \\
\hline
${s1}$ & \multirow{4}{*}{7333.8}
 & \textbf{(i)}  & \textbf{7332.2 $\pm$ 0.5} &  \textbf{1.6 $\pm$ 0.5} 
 &  \textbf{2.9 $\pm$ 0.4} \\
&  & \textbf{(ii)} & \textbf{7333.1 $\pm$ 0.2} &  \textbf{0.7 $\pm$ 0.2} 
&  \textbf{4.5 $\pm$ 0.6}  \\
& & \textbf{(iii)} & \textbf{7333.7 $\pm$ 0.03} &  \textbf{0.1 $\pm$ 0.03 }
 & \textbf{ 12.6 $\pm$ 1.9} \\
& & (iv)  &  7333.79  $\pm$ 0.01&  0.01 $\pm$ 0.01 &  32.7 $\pm$ 4.9 \\
 \hline 
 ${s2}$ & \multirow{4}{*}{ 7527.7 } 
 & (i)  &  7522.4 $\pm$ 2.7 &  5.3 $\pm$ 2.7
  &  1.6 $\pm $ 0.4 \\
&   & \textbf{(ii)} & \textbf{7525.9 $\pm$ 0.9} &  \textbf{1.8 $\pm$ 0.9}
 &  \textbf{2.7 $\pm$ 0.7} \\
&   & \textbf{(iii)} & \textbf{7527.5 $\pm$ 0.1} &  \textbf{0.2 $\pm$ 0.1}
 &  \textbf{9.8 $\pm$ 2.4} \\
&  & (iv)   & 7527.68 $\pm$ 0.01 & 0.02  $\pm$ 0.01 & 28.3  $\pm$ 7.1\\
 \hline 
 ${t}$ & \multirow{4}{*}{ 7333.8 } 
  & (i) & 6315.8  $\pm$ 305.4 &  1018  $\pm$ 305.4 &  0.1 $\pm$ 0.02  \\
&   & (ii) & 7310.9  $\pm$ 6.9 &  22.9 $\pm$ 6.9 &  0.8  $\pm$ 0.2\\
&  & \textbf{(iii)} & \textbf{7333.4 $\pm$ 0.1} &  \textbf{0.4 $\pm$ 0.1}
 &  \textbf{6.1 $\pm$ 0.9} \\
&  & (iv)  & 7333.77  $\pm$ 0.1 &  0.03 $\pm$ 0.01 &  21.6 $\pm$ 3.2 \\
 \hline  
 ${d1}$ & \multirow{4}{*}{ 7437.5 } 
  & (i) & 6524.5 $\pm$ 456.5 &  913 $\pm$ 456.5 &  0.1 $\pm$ 0.03  \\
&  & (ii)  & 7417  $\pm$ 10.3 &  20.5 $\pm$ 10.3 &  0.8 $\pm$ 0.2\\
&   & \textbf{(iii)} & \textbf{7437.2 $\pm$ 0.2} &  \textbf{0.3 $\pm$ 0.2} 
&  \textbf{6.3 $\pm$ 1.6}  \\
&  & (iv)   & 7437.47 $\pm$ 0.1 &  0.03 $\pm$ 0.1 &  22.4 $\pm$ 5.6 \\
 \hline 
${d2}$ & \multirow{4}{*}{ 7424 } 
   & (i)   &6419.9  $\pm$ 502.1 &  1004.1 $\pm$ 502.1 &  0.1  $\pm$ 0.03\\
&   & (ii)  & 7401.4 $\pm$ 11.3 &  22.6 $\pm$ 11.3 &  0.8 $\pm$ 0.2  \\
&    & \textbf{(iii)} & \textbf{7423.6 $\pm$ 0.2} &  \textbf{0.4 $\pm$ 0.2}
 &  \textbf{6.1 $\pm$ 1.5} \\
&   & (iv)  & 7423.7  $\pm$ 0.1 & 0.03  $\pm$ 0.1 &  21.7 $\pm$ 5.4 \\
 \hline   \hline 
\label{table6}
 \end{longtable}

\begin{longtable}{cccccc}
 \caption{ Same as in table \ref{table2} for $D^{*} B^{*}({}^5 S_2)_{\xi}$ system, 
 with ($D2=0.00001 \; \mathrm{MeV}^{-2}; E2=0.00002\; \mathrm{MeV}^{-2}$).  } \\
\hline
$\xi$ & $M_{Th}$ (MeV) & $(D_1, E_1)$ & $M$ (MeV) & B.E. (MeV) & $a_s$ (fm) \\
\hline
${s1}$ & \multirow{4}{*}{7333.8}
& (i)  & -  &  -  &  - \\
&  & (ii) & 7317.9 $\pm$ 4.8 &  15.9 $\pm$ 4.8  &  0.9  $\pm$ 0.1 \\
&   & \textbf{(iii)} & \textbf{7333.6 $\pm$ 0.1} & \textbf{0.2 $\pm$ 0.1}
 & \textbf{9.1 $\pm$ 1.4} \\
& & (iv)  &  7333.78  $\pm$ 0.01 &  0.02  $\pm$ 0.01 &  29.2  $\pm$ 4.3 \\
 \hline 
 ${s2}$ & \multirow{4}{*}{ 7527.7 } 
 & (i) &  -  &  - &  - \\
&    & (ii) & 7513.6  $\pm$ 7 &  14.1 $\pm$ 7 &  1  $\pm$ 0.2\\
&   & \textbf{(iii)} & \textbf{7527.5 $\pm$ 0.1}  &  \textbf{0.2 $\pm$ 0.1}
  &  \textbf{8 $\pm$ 2} \\
&  & (iv)   & 7527.68 $\pm$ 0.01  &  0.02  $\pm$ 0.01 &  26.5 $\pm$ 6.6 \\
 \hline 
 ${t}$ & \multirow{4}{*}{ 7333.8 } 
  & (i) & 7146.8  $\pm$ 56 &  187 $\pm$ 56  &  0.3 $\pm$ 0.1 \\
&   & (ii) & 7317.9  $\pm$ 4.8 & 15.9  $\pm$ 4.8 &  0.9 $\pm$ 0.1 \\
&   & \textbf{(iii)} & \textbf{7333.4 $\pm$ 0.1}  &  \textbf{0.5 $\pm$ 0.1}
  &  \textbf{6.3 $\pm$ 0.9}\\
&  & (iv)  & 7333.77  $\pm$ 0.01 &  0.03  $\pm$ 0.01 &  21.7 $\pm$ 3.3 \\
 \hline  
 ${d1}$ & \multirow{4}{*}{ 7437.5 } 
  & (i) & 7269.8  $\pm$ 83.8 &  167.7 $\pm$ 83.8  &  0.3  $\pm$ 0.1\\
&    & (ii)  & 7423.2 $\pm$ 7.1  &  14.3 $\pm$ 7.1  & 1 $\pm$ 0.2\\
&    & \textbf{(iii)} & \textbf{7437.2 $\pm$ 0.2}  &  \textbf{0.3 $\pm$ 0.2}
  & \textbf{6.5 $\pm$ 1.6}  \\
&  & (iv)   & 7437.47 $\pm$ 0.01  &  0.03  $\pm$ 0.01 &  22.5  $\pm$ 5.6\\
 \hline 
${d2}$ & \multirow{4}{*}{ 7424 } 
   & (i)   & 7239.6  $\pm$ 92.2 &  184.4 $\pm$  92.2 &  0.3  $\pm$ 0.1\\
&     & (ii)  & 7408.3 $\pm$ 7.8   &  15.7 $\pm$ 7.8 & 0.9 $\pm$ 0.2 \\
&   & \textbf{(iii)} & \textbf{7423.7 $\pm$ 0.2}  &  \textbf{0.3 $\pm$ 0.2}
  &  \textbf{6.3 $\pm$ 1.6} \\
&   & (iv)  & 7423.97 $\pm$ 0.01  &  0.03 $\pm$ 0.01  &  21.8 $\pm$ 5.5 \\
 \hline   \hline 
\label{table7}
 \end{longtable}

\subsection{Quantities for the $D^{(*)}B^{(*)} $ systems for fixed values of 
$(D_1, E_1,D_2, E_2)$ obtained from inputs}
\label{AppC}

\begin{longtable}{cccccc}
\caption{Relevant quantities for the $D^{(*)}B^{(*)} $ systems. The 
 parameters $(D_1, E_1,D_2, E_2)$ are chosen
in the situation: $ D_1=-0.0003973 \; \mathrm{MeV}^{-2}, \;  D_2 = 0.00027 
 \; \mathrm{MeV}^{-2}, \; D_2 = -0.00017037\; \mathrm{MeV}^{-2}, \; E_2= 
 -0.0001351\; \mathrm{MeV}^{-2}$. 
 $M_{Th}$, $M$, B.E., and $a_s$ mean threshold, mass, binding energy and 
scattering length of the respective state. } \\
\hline
State &  $ \xi $ & $M_{Th}$ (MeV) & $M$ (MeV) & B. E. (MeV) & $a_s$ (fm) \\
 \hline 
 \multirow{5}{*}{  $DB({}^1 S_0)$  }  
  & $s1$ & 7146.6 & 7146.58  $\pm$ 0.01 &  0.02 $\pm$ 0.01  &  28.5 $\pm$ 6.9 \\
  & $s2$ & 7335.3 & 7335.3   & -  &  - \\
  & $t$ & 7146.6 & 7146 $\pm$ 0.3 &  0.6 $\pm$ 0.3  &  4.8 $\pm$ 0.9 \\
  & $d1$ & 7247.9 & 7247.3  $\pm$ 0.3 &  0.6 $\pm$ 0.3 & 5 $\pm$ 1.4 \\
  & $d2$ & 7234   & 7233.4  $\pm$ 0.4 &  0.6 $\pm$ 0.4  &  4.8 $\pm$ 1.3 \\
 \hline 
 \multirow{5}{*}{  $D^{*}B({}^3 S_1)$  }  
  & $s1$ & 7288    & 7287.98  $\pm$ 0.01 &  0.02 $\pm$ 0.01  &  30 $\pm$ 4.9 \\
  & $s2$ & 7479.1  & 7479.06 $\pm$ 0.02  & 0.04  $\pm$ 0.02 & 18.3 $\pm$  4.7 \\
  & $t$  & 7288    & 7287.5 $\pm$ 0.2 &  0.5 $\pm$ 0.2  &  5 $\pm$ 1 \\
  & $d1$ & 7391.7  & 7391.2  $\pm$ 0.3 &  0.5 $\pm$ 0.3  & 5.2 $\pm$ 1.5\\
  & $d2$ & 7375.4  & 7374.9  $\pm$ 0.3 &  0.5 $\pm$ 0.3  &  5.1 $\pm$ 1.4 \\
 \hline 
  \multirow{5}{*}{  $D B^{*}({}^3 S_1)$  }  
  & $s1$ & 7192.4 & 7192.38  $\pm$ 0.01 &  0.02 $\pm$ 0.01  &  28.5 $\pm$ 4.6 \\
  & $s2$ & 7383.9 & 7383.8 $\pm$ 0.02  & 0.02  $\pm$ 0.02 & 17.4 $\pm$ 4.4 \\
  & $t$  & 7192.4 & 7191.8 $\pm$ 0.3 &  0.6 $\pm$ 0.3  &  4.8 $\pm$ 0.9 \\
  & $d1$ & 7293.7 & 7293.1  $\pm$ 0.3 &  0.6 $\pm$ 0.3  & 5 $\pm$ 1.4 \\
  & $d2$ & 7282.6 & 7282  $\pm$ 0.3 &  0.6 $\pm$ 0.3  &  4.8 $\pm$ 1.4 \\
 \hline 
\multirow{5}{*}{  $D^{*} B^{*}({}^1 S_0)$  }  
  & $s1$ & 7333.8 & 7325.6  $\pm$ 33.5 &  8.2 $\pm$ 33.5  &  1.3 $\pm$ 2.6 \\
  & $s2$ & 7527.7 & 7520.6 $\pm$ 14.2  & 7.1  $\pm$ 14.2 & 1.4 $\pm$ 1.4 \\
  & $t$  & 7333.8 & 7325.9 $\pm$ 11.2 &  7.9 $\pm$ 11.2  &  1.3 $\pm$ 0.9 \\
  & $d1$ & 7437.5 & 7430.4  $\pm$ 10.4 &  7.1 $\pm$ 10.4  & 1.4 $\pm$ 1 \\
  & $d2$ & 7424   & 7416.2  $\pm$ 11.4 &  7.8 $\pm$ 11.4  &  1.3 $\pm$ 1 \\
 \hline
 \multirow{5}{*}{  $D^{*} B^{*}({}^3 S_1)$  }  
  & $s1$ & 7333.8 & 7333.7  $\pm$ 0.03 &  0.1 $\pm$ 0.02  &  15.7 $\pm$ 3.5 \\
  & $s2$ & 7527.7 & 7527.6 $\pm$ 0.1  & 0.1  $\pm$ 0.1 & 9.8 $\pm$ 2.8 \\
  & $t$  & 7333.8 & 7332.4 $\pm$ 0.9 &  1.4 $\pm$ 0.9  &  3.1 $\pm$ 1 \\
  & $d1$ & 7437.5 & 7436.3  $\pm$ 0.9 &  1.2 $\pm$ 0.9  & 3.3 $\pm$ 1.3 \\
  & $d2$ & 7424   & 7422.7  $\pm$ 0.8 &  1.3 $\pm$ 0.8  &  3.2 $\pm$ 0.9 \\
 \hline
 \multirow{5}{*}{  $D^{*} B^{*}({}^5 S_2)$  }  
  & $s1$ & 7333.8 & 7333.79  $\pm$ 0.01 &  0.01 $\pm$ 0.01  &  44.5 $\pm$ 7.2 \\
  & $s2$ & 7527.7 & 7527.68 $\pm$ 0.01  & 0.2  $\pm$ 0.01 & 26.8 $\pm$ 6.8 \\
  & $t$  & 7333.8 & 7333.5 $\pm$ 0.1 &  0.3 $\pm$ 0.1  & 6.9 $\pm$ 1.4 \\
  & $d1$ & 7437.5 & 7437.2  $\pm$ 0.1 &  0.3 $\pm$ 0.1  & 7.2 $\pm$ 2 \\
  & $d2$ & 7424   & 7423.7  $\pm$ 0.2 &  0.3 $\pm$ 0.2  &  6.9 $\pm$ 2 \\
 \hline 
 \hline 
\label{table8}
 \end{longtable}


\end{document}